\documentclass[twocolumn,showpacs,preprintnumbers,amsmath,amssymb,prb,superscriptaddress]{revtex4-1}
\usepackage{amsfonts,amsmath,amssymb}
\usepackage{bbm}
\usepackage[english]{babel}
\usepackage[T1]{fontenc}
\usepackage{times}
\usepackage{mathrsfs}
\usepackage{graphicx}
\usepackage{dcolumn}
\usepackage{bm}
\usepackage[colorlinks,bookmarks=true,citecolor=blue,linkcolor=red,urlcolor=blue]{hyperref}
\usepackage[tight, FIGTOPCAP, normal, raggedright, nooneline]{subfigure}
\usepackage{hyperref}
\usepackage{bbm}

\renewcommand{\vec}[1]{\mathbf{#1}}

\newcommand{\vM}{\vec M}

\newcommand{\vsigma}{\mbox{\boldmath $\sigma$}}

\begin{document}

\title{Interference
effects induced by a precessing easy-plane magnet coupled to a helical
edge state}

\author{Kevin A.~Madsen}
\email{kevin.madsen@fu-berlin.de}
\affiliation{Dahlem Center for Complex Quantum Systems and Institut f\"ur Physik, Freie Universit\"at Berlin, Arnimallee 14, D-14195 Berlin, Germany}
\author{Piet W.~Brouwer}
\affiliation{Dahlem Center for Complex Quantum Systems and Institut f\"ur Physik, Freie Universit\"at Berlin, Arnimallee 14, D-14195 Berlin, Germany}
\author{Patrik Recher}
\affiliation{Institut f\"ur Mathematische Physik, Technische Universit\"at Braunschweig, D-38106 Braunschweig, Germany}
\affiliation{Laboratory for Emerging Nanometrology Braunschweig, D-38106 Braunschweig, Germany}
\author{Peter G.~Silvestrov}
\affiliation{Institut f\"ur Mathematische Physik, Technische Universit\"at Braunschweig, D-38106 Braunschweig, Germany}

\date{\today}

\begin{abstract}

The interaction of a magnetic insulator with the helical electronic edge of a two-dimensional topological insulator has been shown to lead to many interesting phenomena. One of these is that for a suitable orientation of the magnetic anisotropy axis, the exchange coupling to an easy-plane magnet has no effect on DC electrical transport through a helical edge, despite the fact that it opens a gap in the spectrum of the helical edge [Meng {\em et al.}, Phys.\ Rev.\ B {\bf 90}, 205403 (2014)]. Here, we theoretically consider such a magnet embedded in an interferometer, consisting of a pair of helical edge states connected by two tunneling contacts, at which electrons can tunnel between the two edges. Using a scattering matrix approach, we show that the presence of the magnet in one of the interferometer arms gives rise to AC currents in response to an applied DC voltage. On the other hand, the DC Aharonov-Bohm effect is absent at zero temperature and small DC voltages, and only appears if the applied voltage or the temperature exceeds the magnet-induced excitation gap. 
\end{abstract}

\maketitle

\section{Introduction}\label{intro}

Since backscattering of electrons in the helical edge of a two-dimensional topological insulator is forbidden by time-reversal symmetry, breaking time-reversal symmetry by an applied magnetic field or by coupling of the helical edge to the exchange field of a magnet or a magnetic impurity is the only mechanism by which electrons in a helical edge can be backscattered.\cite{kane2005,bernevig2006,wu2006,xu2006,koenig2007,du2015} The purposeful coupling of the helical edge to magnetic insulators has been shown to result in fascinating properties, such as the appearance of Majorana zero modes at the boundary between segments with a magnet-induced gap and with proximity-induced superconductivity,\cite{fu2008} various thermoelectric effects,\cite{bas2018,gresta2019} or the possibility to convert electrical energy to mechanical motion in an adiabatic quantum motor.\cite{arrachea2015,bruch2018} Magnetic impurities exchange-coupled to the helical edge states exhibit characteristic Kondo effects\cite{maciejko2009,tanaka2011,altshuler2013,posske2013} and electrically controlled dynamics of the impurity spin due to backscattering of helical edge state electrons.\cite{probst2015,locane2017}

The application of a magnetic field or the exchange coupling of a helical edge to a magnetic insulator opens a gap in the spectrum of the helical edge if the direction of the (exchange) field is not collinear with the spin quantization axis of the helical edge states. In this context, it can be seen as a surprise that an electric current carried by a helical edge is transmitted perfectly across a region coupled to a magnetic insulator, if the magnet has an easy-plane anisotropy with easy plane perpendicular to the quantization axis of the helical edge states.\cite{meng2014,silvestrov2016} The electrical current flows despite the presence of an excitation gap in the spectrum of the helical edge, in such a way that the electrical current ``lost'' by the backscattering of electronic quasiparticles at the gapped region is compensated by the flow of a dissipationless spin current carried by the precessing magnetization and facilitated by charge-to-spin and spin-to-charge conversion at the magnet interface. The charge-to-spin conversion at the magnet interface is perfect because of the helical nature of the edge;\cite{qi2008} the absence of losses for the spin transport through the magnet is a manifestation of ``superfluid'' spin transport in easy-plane magnets.\cite{koenig2001,nogueira2004,sonin2010,takei2014,takei2015,skarsvaag2015} Since the electrical current is carried by a collective mode of the magnetization, it takes place without shot noise and with strongly suppressed thermal noise at finite frequencies.\cite{silvestrov2016} The coupling of a helical edge to a magnetic insulator has also been proposed as a method for nondissipative current-driven magnetization precession.\cite{arrachea2015,xiao2021}

The perfect compensation of the current backscattered from the magnet-induced gap leaves the open question: How can the presence of the easy-plane magnet be detected, if the current through the edge is not influenced by it? In order to answer this question we theoretically consider an Aharonov-Bohm interferometer\cite{aharonov1959} consisting of two tunnel-coupled helical edge modes, one of which is covered by an easy-plane magnet like in Refs.\ \onlinecite{meng2014,silvestrov2016}. 

We predict the two main characteristic features: First, we show that this interferometer setup, if subjected to a constant (DC) voltage bias, responds with a time-dependent (AC) current component (in addition to a large DC response), due to the coupling to the precessing magnetization. Depending on system parameters, the AC currents occur at the precession frequency $\omega_{\rm M}$ of the magnet or, additionally, at twice that frequency. For comparison: No AC current response exists in the setup analyzed in Refs.\ \onlinecite{meng2014,silvestrov2016}. Second, the Aharonov-Bohm oscillations in the DC current, usually seen as a signature of coherent quasiparticle transport, are exponentially suppressed for small bias voltages and temperatures if the Fermi level is in the magnet-induced gap. The suppression of Aharonov-Bohm oscillations is consistent with the existence of an excitation gap in the spectrum where the helical edge is in contact with the magnet. Despite the absence of a conventional Aharonov-Bohm effect, in our geometry the AC currents can be seen as a manifestation of coherence: they result from the interference of electrons scattered at the tunnel contacts and electrons backscattered from the magnet interface, where they change their energy by the amount $\hbar \omega_{\rm M}$.

The appearance of AC currents in response to a DC voltage bias is remotely reminiscent of the Josephson effect, where applying a constant voltage bias to a superconducting tunnel junction leads to time-dependent currents.\cite{josephson1962} It may be seen as another manifestation of ``spin superfluidity'' in easy-plane ferromagnets.\cite{koenig2001,nogueira2004,sonin2010,takei2014,takei2015,skarsvaag2015}

Electron interferometers with helical edge states, but without exchange coupling to a magnetic insulator, have been investigated theoretically in the literature. The characteristic temperature and interaction-induced dephasing\cite{virtanen2011} of charge and spin excitations were identified, as well as controllable spin properties.\cite{dolcini2011,dolcetto2016,maciejko2010} On the experimental side, quantum point contacts between two helical edge channels have been realized recently in HgTe-based quantum wells and DC transport through the constriction has been measured.\cite{strunz2020} The possibility to produce AC currents if these systems are exchange-coupled to magnetic insulators provides a promising novel route for interference-based quantum devices.

This work is organized as follows: In Sec.\ \ref{setup} the interferometer setup is described in detail and the scattering-matrix approach is introduced, extending the method of Ref.\ \onlinecite{silvestrov2016}. In Secs.\ \ref{sec:open}, \ref{sec:closed}, and \ref{sec:full} we present calculations of the current in response to a DC bias in one of the interferometer arms. Section \ref{ssec:opencurrents} addresses a simplified setup in which one of the two tunnel contacts is ``open,'' which has AC currents at frequency $\omega_{\rm M}$ only, but allows all calculations to be performed analytically. This analytical result allows us to elucidate the difference between the
DC current pumped through the magnet and the interference AC current
created upon reflection from the rotating magnetization
vector. Section \ref{ssec:cl-vs-qu} provides additional details on the interpretation of these results. Section \ref{sec:closed} considers the special case that one of the two tunnel contacts is ``closed,'' which shows the full phenomenology of AC currents at frequencies $\omega_{\rm M}$ and $2 \omega_{\rm M}$ and still admits a partially analytical treatment. Section \ref{sec:full} contains our results for an interferometer with a generic choice of parameters. In Section \ref{sec:finiteTAB} the anomalous temperature dependence of DC Aharonov-Bohm current contributions is discussed. We conclude in Sec.\ \ref{sec:conclusion}.

\section{Model}\label{setup}

\begin{figure}
	\centering
	\includegraphics[width=.21\textwidth]{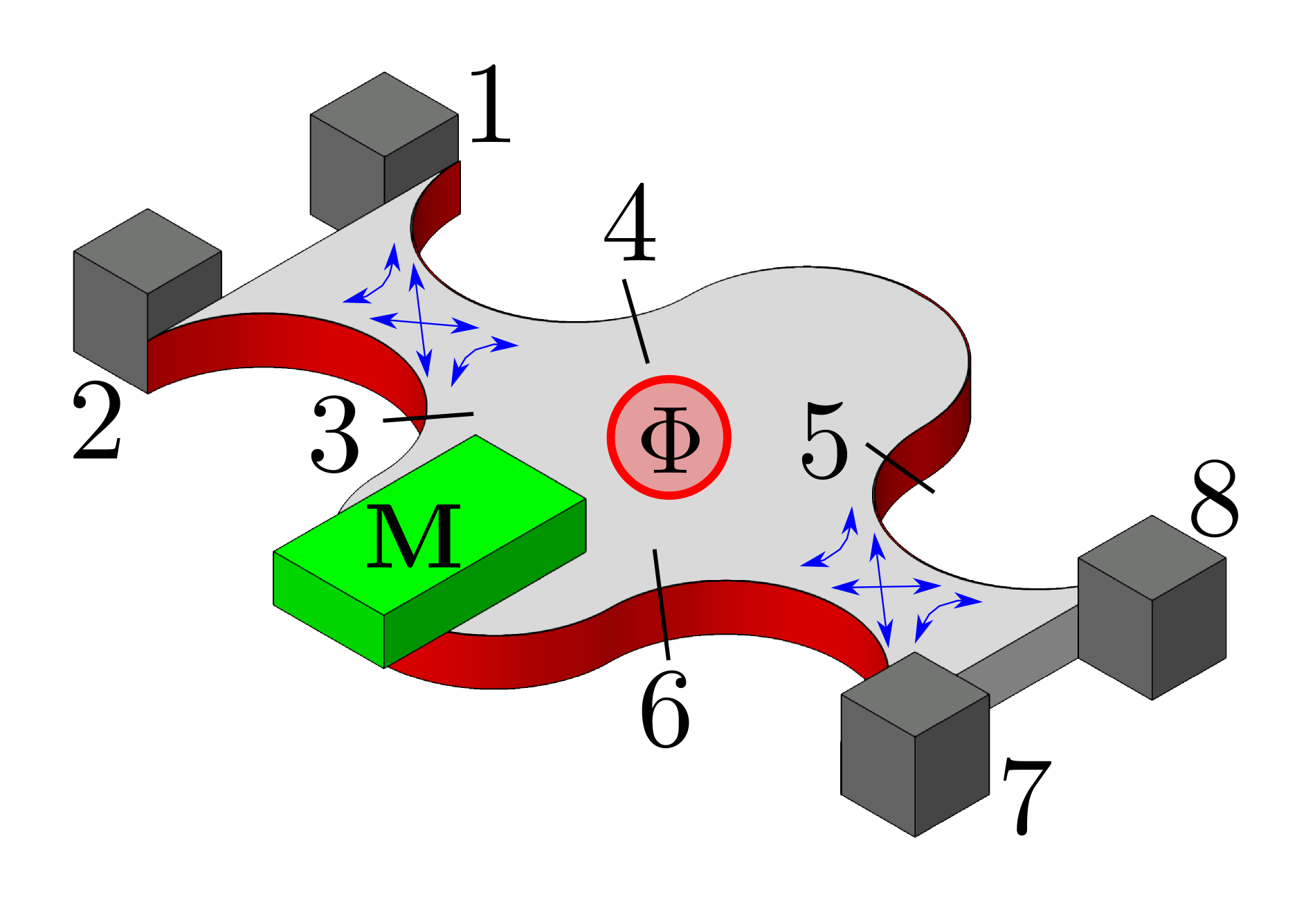}
	\includegraphics[width=.26\textwidth]{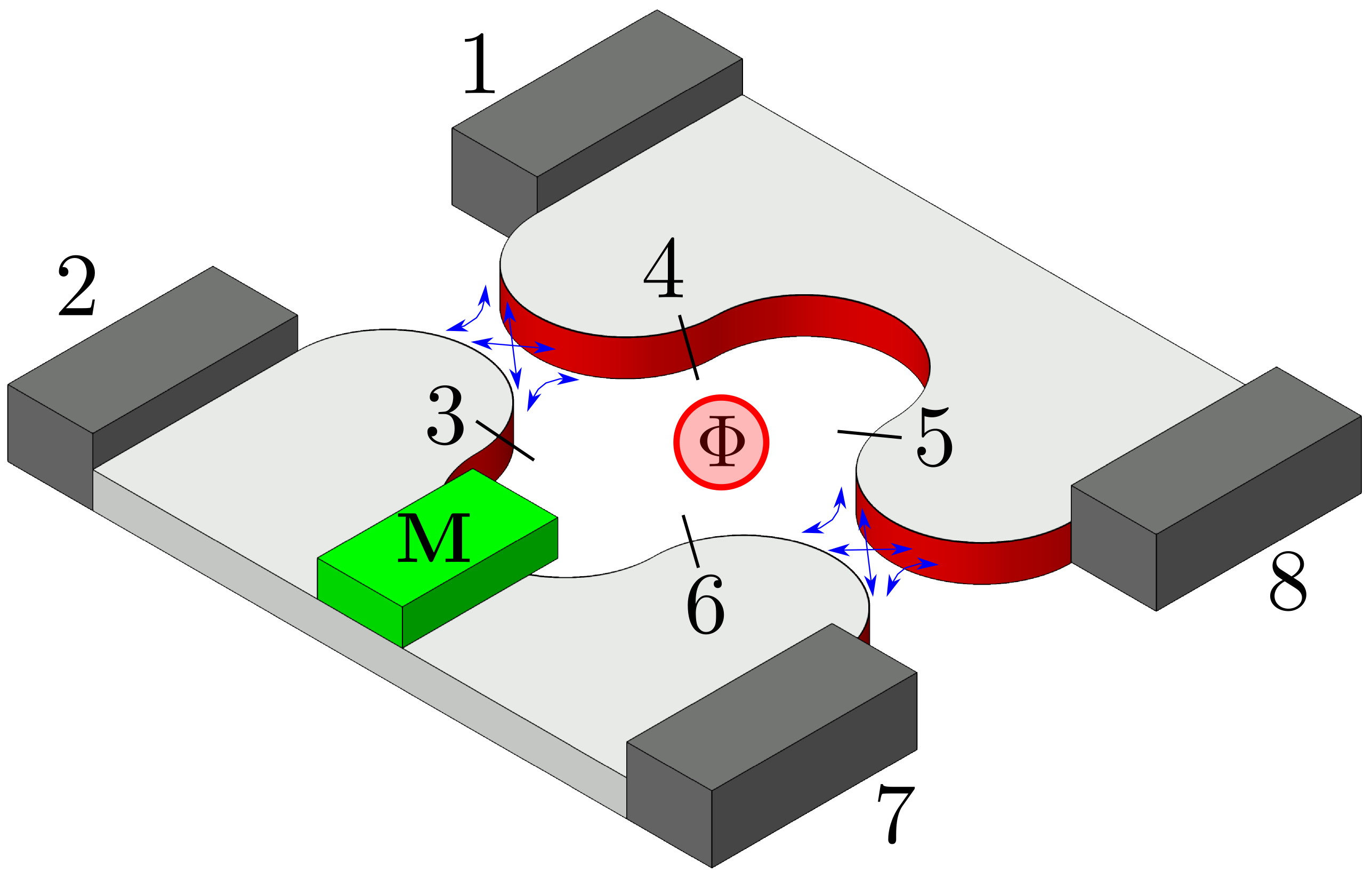}
	\caption{Schematic picture of the interferometer considered here. The interferometer can be realized by bringing helical edge channels of the same quantum spin Hall insulator sufficiently close together that they can make electrical contact (left), or by bringing the edge channels of different quantum spin Hall insulators into contact (right). Both panels show the location of eight reference points $i=1,2,\ldots,8$ used for the calculations in the main text. The ``positive'' current direction for $i=1,2,3,4$ and for $i=5,6,7,8$ is to the right and to the left, respectively. The interferometer is threaded by a magnetic flux $\Phi$.}
	\label{fig:setup}
\end{figure}

We consider an interferometer built from two opposing helical edge modes of a quantum spin Hall insulator. One of the arms of the interferometer is partially covered by an insulating magnet. Following Refs.\ \onlinecite{meng2014,silvestrov2016} we consider a magnet with an easy plane anisotropy, such that the easy-plane is oriented perpendicular to the spin quantization axis of the helical edge mode. Establishing contact between the two edge channels on both sides of the magnet then results in the typical interferometer geometry (see Fig.\ \ref{fig:setup}). The interferometer is connected to four ideal leads, as shown schematically in the figure.

In practice, such a geometry may be realized by taking the helical edge modes on the two sides of the same quantum spin Hall insulator [see Fig.\ \ref{fig:setup} (left)]. In this case, only part of the insulator is covered by the magnetic insulator. In this geometry, a ``point contact'' between the two sets of edge modes can be achieved, e.g., by locally reducing the insulator width by lithographic methods or by electrostatic gating of the device. Alternatively, the helical states can be edges of different spin Hall insulators, and contacting the to edge modes is achieved by bringing the edges close together [see Fig.\ \ref{fig:setup} (right)].

A magnetic flux $\Phi$ is threaded through the area enclosed by the interfering edge modes. Apart from the magnetic flux and the coupling to the magnetic insulator, time-reversal symmetry is unbroken, so that backscattering into the same edge mode is forbidden everywhere in the device (except in the vicinity of the magnetic insulator). 

The magnet is assumed to be small enough, that it may be described by a single moment $\vM$. We choose the (spin) coordinate axes such that the easy plane is the $xy$ plane and the quantization axis for the spin in the helical edge mode is the $z$ axis. With this choice, the magnet and its interaction with the helical edge state are described by the second-quantized Hamiltonian \cite{meng2014,silvestrov2016}
\begin{align}
  \label{eq:H}
  \hat H =&\, \int dx \hat \psi(x)^{\dagger}
  [-i \hbar v_{\rm F} \partial_x \sigma_z + h(x) \vsigma \cdot \vM] \hat \psi(x)
  \nonumber \\ &\, + \frac{D}{2} M_z^2,
\end{align}
where $v_{\rm F}$ is the Fermi velocity, $D$ the strength of the easy-plane anisotropy, $\sigma_{x,y,z}$ are the Pauli matrices, $h(x)$ is a function that describes the exchange coupling between the magnetic moment $\vM$ and the spin of electrons at the helical edge state, and $\hat \psi(x) = \left[ \hat \psi_{\uparrow}(x),\hat \psi_{\downarrow}(x) \right]^{\rm T}$ is a two-component spinor describing electrons in the helical edge. Away from the magnet, the exchange coupling $h(x) \to 0$. The assumption that the easy plane is perpendicular to the spin quantization axis is generic for a thin magnetic film \cite{ohandley2000} exchange coupled to the helical edge modes of a quantum spin Hall material, such as a HgTe quantum well.\cite{brune2012} The application of a finite bias across the magnet causes the magnetization to cant out of the $xy$ plane,\cite{meng2014,silvestrov2016} which sets the magnet in a precessional motion with frequency 
\begin{equation}
  \omega_{\rm M} = D M_z. \label{eq:1}
\end{equation}

\begin{figure}
	\centering
	\includegraphics[width=.45\textwidth]{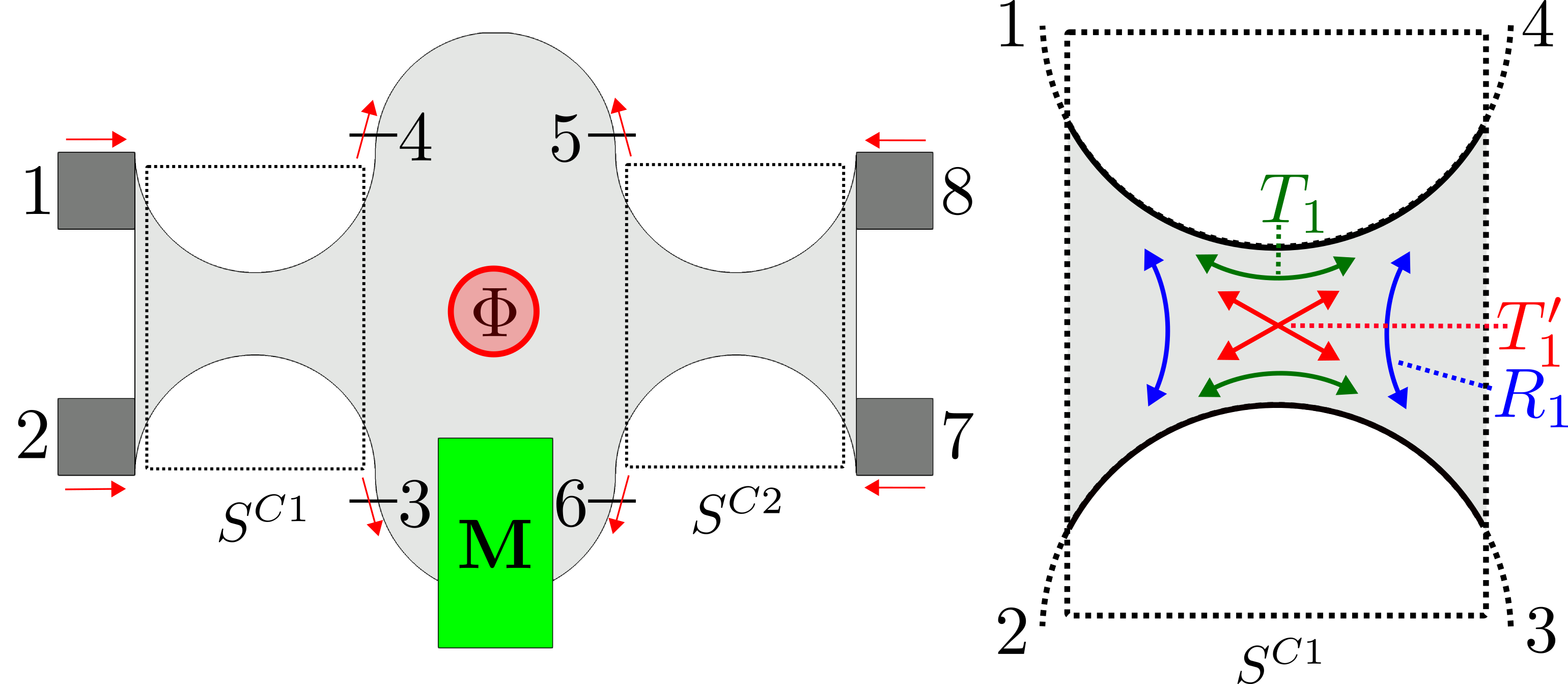}~~

	\caption{In the theoretical description, currents are calculated for eight reference points $i=1,2,\ldots,8$, as shown in the left panel. The ``positive'' current direction for $i=1,2,3,4$ and for $i=5,6,7,8$ is to the right and to the left, respectively. The two tunneling point contacts between the opposing helical edge states are described by scattering matrices $S^{({\rm C}1)}$ and $S^{({\rm C}2)}$. The definitions of the transmission coefficients $T_1$ and $T_1'$ and the reflection coefficient $R_1$ for the left tunneling point contact are shown in the right panel.\label{fig:bigpicture}}
\end{figure}

We describe coherent transport through the interferometer using scattering theory. To this end, we mark eight reference positions labeled $i=1,2,\ldots, 8$ in the device (see Fig.\ \ref{fig:setup}). The reference positions are chosen in close proximity to the point-contact region, where scattering between the helical edge states is possible. At each reference position $i$ we consider creation and annihilation operators $\hat a_{i,\pm}^{\dagger}(\varepsilon)$ and $\hat a_{i,\pm}(\varepsilon)$ for an electron in a (particle-flux normalized) scattering state at energy $\varepsilon$, moving in the same ($+$) or opposite ($-$) direction as the reference arrows in Fig.\ \ref{fig:bigpicture}.
The operators $\hat a_{i\pm}(\varepsilon)$ and the corresponding creation operators $\hat a^{\dagger}_{i\pm}(\varepsilon)$ are related to the current ${\cal I}_i(\omega)$ at reference position $i$ and at frequency $\omega$ as\cite{buettiker1992,blanter2000}
\begin{align}
  \label{eq:Iomega}
  {\cal I}_i(\omega) =&\, \frac{e}{h} 
  \int d\varepsilon
  \\ \nonumber &\, \mbox{} \times
  \left[\overline{\hat a_{i+}^{\dagger}(\varepsilon) \hat a_{i+}(\varepsilon + \hbar \omega)} -
   \overline{\hat a_{i-}^{\dagger}(\varepsilon) \hat a_{i-}(\varepsilon + \hbar \omega)} \right].
\end{align}
Here, $\overline{\cdots}$ denotes the expectation value.
Electrons coming in from the four ideal leads (corresponding to the reference positions $i=1,2,7,8$) are in thermal equilibrium at temperature $T_i$ and chemical potential $e V_i$,
\begin{align}
  \overline{\hat a_{i+}^{\dagger}(\varepsilon) \hat a_{i+}(\varepsilon')}
  =&\, f_{i}(\varepsilon) \delta(\varepsilon-\varepsilon'),\ \ i=1,2,7,8,
\end{align}
where $f_i(\varepsilon) = [1 + e^{(\varepsilon - e V_i)/k_{\rm B} T_i}]^{-1}$ is the Fermi-Dirac distribution function.

Scattering is elastic everywhere in the device, except at the magnet, where electrons can absorb or emit an energy quantum $\hbar \omega_{\rm M}$ upon reflection.\cite{silvestrov2016}
Elastic scattering from the two point-contact regions is described by $4 \times 4$ scattering matrices $S^{({\rm C}1)}$ and $S^{({\rm C}2)}$. We assume that the point-contact regions are small enough that $S^{({\rm C}1)}$ and $S^{({\rm C}2)}$ may be taken independent of the energy $\varepsilon$,
\begin{align}
  \begin{pmatrix}
  \hat a_{1-}(\varepsilon) \\ 
  \hat a_{2-}(\varepsilon) \\ 
  \hat a_{3+}(\varepsilon) \\ 
  \hat a_{4+} (\varepsilon)
  \end{pmatrix}
  =&\,
  S^{({\rm C}1)}
   \begin{pmatrix}
  \hat a_{1+}(\varepsilon) \\ 
  \hat a_{2+}(\varepsilon) \\ 
  \hat a_{3-}(\varepsilon) \\ 
  \hat a_{4-} (\varepsilon)
  \end{pmatrix}, \nonumber \\
  \begin{pmatrix}
  \hat a_{5+}(\varepsilon) \\ 
  \hat a_{6+}(\varepsilon) \\ 
  \hat a_{7-}(\varepsilon) \\ 
  \hat a_{8-} (\varepsilon)
  \end{pmatrix}
  =&\,
  S^{({\rm C}2)}
   \begin{pmatrix}
  \hat a_{5-}(\varepsilon) \\ 
  \hat a_{6-}(\varepsilon) \\ 
  \hat a_{7+}(\varepsilon) \\ 
  \hat a_{8+}(\varepsilon)
  \end{pmatrix}. 
  \label{eq:4}
\end{align}
Time-reversal symmetry imposes the antisymmetry constraints
\begin{equation}
  S^{({\rm C}j)} = -(S^{({\rm C}j)})^{\rm T}.
\end{equation}
Absorbing eventual phase factors in the definitions of the operators $\hat a_{i \pm}$, this implies that without loss of generality these matrices can be parametrized as
\begin{equation}
  S^{({\rm C}j)} =
  \begin{pmatrix}
    0 & \sqrt{R_j} & -i \sqrt{T_{j}'} & - \sqrt{T_{j}} \\
  - \sqrt{R_j} & 0 & - \sqrt{T_{j}} & -i \sqrt{T_{j}'} \\
  i \sqrt{T_{j}'} & \sqrt{T_{j}} & 0 & \sqrt{R_j} \\
  \sqrt{T_{j}} & i \sqrt{T_{j}'} & -\sqrt{R_j} & 0
  \end{pmatrix} ,
  \label{eq:SCj}
\end{equation}
with $R_j + T_j + T_j' = 1$ and $j=1,2$. As can be seen in Fig.\ \ref{fig:bigpicture}, $R_j$, $T_j$ and $T_j'$ are the probabilities for reflection via point-contact $j$ (changing the edge), transmission along the same edge, and transmission through the point-contact $j$ (changing the edge), respectively. Note that $R_j$ and $T_j$ describe both spin-conserving processes, whereas the process due to $T_j'$ flips the electron spin. The latter process is possible even if time-reversal symmetry is conserved but needs a breaking of inversion symmetry \cite{liu2008,rothe2010,sternativo2014}. Tunable spin-flip processes due to Rashba spin-orbit coupling become possible by an electric field \cite{rothe2010,orth2013,sternativo2014} that could be induced locally by gates. We will see in the upcoming sections that a characteristic AC part of the current becomes possible due to interference between two scattering paths: one that includes reflection at the magnet, and another that does not. These interference contributions necessarily involve processes with amplitude $\sqrt{T_j'}$. 

Since there is no backscattering for propagation along the helical edge, the operators at the two ends of the upper interferometer arm at reference positions ``4'' and ``5'' are simply related by a phase factor,
\begin{align}
  \hat a_{5-}(\varepsilon) =&\, e^{i k(\varepsilon) L + i \phi} \hat a_{4+}(\varepsilon), \nonumber \\
  \hat a_{4-}(\varepsilon) =&\, e^{i k(\varepsilon) L - i \phi} \hat a_{5+}(\varepsilon).
  \label{eq:5}
\end{align}
Here we have chosen a gauge such that the Aharonov-Bohm (AB) phase shift $\phi = e \Phi/\hbar$ from the magnetic flux $\Phi$ is accumulated for the propagation between reference positions ``4'' and ``5''. Further, $L$ is the length of the interferometer arm and $k(\varepsilon) = k_{\rm F} + \varepsilon/\hbar v_{\rm F}$, with $k_{\rm F}$ the Fermi  wavenumber and $v_{\rm F}$ the Fermi velocity. The energy $\varepsilon$ is measured with respect to the Fermi level. 

Because of the spin-momentum locking in the helical edge modes, a reflection from the magnetic insulator necessarily comes with a spin flip of the edge electrons. As a result, reflection from the magnetic insulator involves an increase or decrease of $M_z$ by one and, hence, the absorption or emission of an energy quantum $\hbar \omega_{\rm M}$ by the reflected electron.\cite{meng2014,silvestrov2016} These processes are described by the relation \cite{silvestrov2016}
\begin{equation}
  \begin{pmatrix}
  \hat a_{3-}(\varepsilon_-) \\
  \hat a_{6-}(\varepsilon_+)
  \end{pmatrix}
  = S^{(\rm M)}(\varepsilon)
  \begin{pmatrix}
  \hat a_{3+}(\varepsilon_+) \\
  \hat a_{6+}(\varepsilon_-)
  \end{pmatrix},
  \label{eq:6}
\end{equation}
where $\varepsilon_{\pm} = \varepsilon \pm \hbar \omega_{\rm M}/2$ and the $2 \times 2$ matrix $S^{(\rm M)}(\varepsilon)$ reads
\begin{equation}
  S^{(\rm M)}(\varepsilon) =
  \begin{pmatrix}
  r_{\rm M}(\varepsilon) \hat m_+ &
  t'_{\rm M}(\varepsilon) \\
  t_{\rm M}(\varepsilon) &
  r'_{\rm M}(\varepsilon) \hat m_-
  \end{pmatrix}.
\end{equation}
Here $r_{\rm M}$, $r'_{\rm M}$, $t_{\rm M}$, and $t'_{\rm M}$ are reflection and transmission amplitudes with a ``frozen'' magnetization of the magnet,\cite{silvestrov2016} and $\hat{m}_{\pm}$ are ladder operators that change $M_z$ by $\pm \hbar$ normalized as $\hat{m}_- \hat{m}_+ = \mathbbm{1}$. The amplitudes $r_{\rm M}$, $r'_{\rm M}$, $t_{\rm M}$, and $t'_{\rm M}$ also include phase shifts accumulated during the propagation between the point contacts and the insulating magnet. Unitarity gives the conditions $|r_{\rm M}(\varepsilon)|^2 = |r'_{\rm M}(\varepsilon)|^2 = 1 - |t_{\rm M}(\varepsilon)|^2 = 1 - |t'_{\rm M}(\varepsilon)|^2$. We will assume that $|r_{\rm M}(\varepsilon)| \to 0$ for energies $\varepsilon$ far above and below the Fermi level. This is consistent with the model (\ref{eq:H}), which has $|r_{\rm M}(\varepsilon)| \to 0$ for $|\varepsilon| \gg \max_x|h(x)||M|$.

Upon reflection off of the magnet, an electron changes its energy by the amount $\pm\hbar \omega_{\rm M}$, where the sign of the change is opposite for reflection from the left and from the right [see Eq.\ (\ref{eq:6})]. Since backscattering in the helical channels away from the magnet is forbidden, the difference of the total numbers of reflections of an electron from the left and from the right sides of the magnet cannot be larger than one, so that an electron can not change its energy by more than $\hbar \omega_{\rm M}$ upon moving through the interferometer.

Taken together, Eqs.\ (\ref{eq:4}), (\ref{eq:5}), and (\ref{eq:6}) give a set of 12 linear equations, which allow one to express all operators $\hat a_{i\pm}$ at the reference positions $i=1,2,\ldots,8$ in terms of the four operators $\hat a_{1+}$, $\hat a_{2+}$, $\hat a_{7+}$, and $\hat a_{8+}$ describing electrons incident from the reservoirs. Since the energy $\varepsilon$ can not change by more than one discrete quantum $\hbar \omega_{\rm M}$, the solution of the set of linear equations (\ref{eq:4}), (\ref{eq:5}), and (\ref{eq:6}) can be cast in the form
\begin{align} \label{eq:Sdef}
  \hat{a}_{j-}(\varepsilon) =&\, \sum_{n = -1}^{1} \sum_{k=1,2,7,8}
  S^{(n)}_{j;k}(\varepsilon)
  \hat a_{k+}(\varepsilon + n \hbar \omega_{\rm M}) ,
\end{align}
where $S^{(n)}_{j;k}(\varepsilon)$ is the ``scattering matrix'' of the device. Since the current ${\cal I}(\omega)$ is bilinear in the creation and annihilation operators [see Eq.\ (\ref{eq:Iomega})], it then follows that at any position in the device the current ${\cal I}_j(\omega)$ can be nonzero for $\omega = 0$, $\omega = \pm \omega_{\rm M}$, or $\pm 2\omega_{\rm M}$ only. This allows us to write
\begin{equation}
  {\cal I}_j(\omega) = \sum_{n=-2}^{2} I_j(n \omega_{\rm M}) \delta(\omega - n \omega_{\rm M}).
\end{equation}
Higher harmonics than $|n|=2$ are not possible, as electrons cannot change their energy by more than $\hbar \omega_{\rm M}$ upon passing through the interferometer device. It follows that the DC current in lead $j$ ($j=1,2,7,8$) is
\begin{align}
  I_j(0) =&\, \frac{e}{h} \int d\varepsilon \sum_{k=1,2,7,8}
  \sum_{n=-1}^{1} |S^{(n)}_{j;k}(\varepsilon)|^2
  \nonumber \\ &\, \mbox{} \times
  \left[ f_j(\varepsilon) - f_{k}(\varepsilon + n \hbar \omega_{\rm M}) \right],
  \label{eq:IdcS}
\end{align}
where we used that $\sum_{k,n} |S^{(n)}_{j;k}(\varepsilon)|^2 = 1$. Similarly, the AC currents at frequency $\omega_{\rm M}$ and $2 \omega_{\rm M}$ read
\begin{align}
  I_j(\omega_{\rm M}) =&\, - \frac{e}{h} \int d\varepsilon \sum_{k=1,2,7,8}
  \sum_{n=0}^{1} S^{(n)}_{j;k}(\varepsilon)^*   \nonumber \\ &\, \mbox{} \times
 S^{(n-1)}_{j;k}(\varepsilon+\hbar \omega_{\rm M}) f_k(\varepsilon+n \hbar \omega_{\rm M}),  \label{eq:IacS1} \\
  I_j(2 \omega_{\rm M}) =&\, - \frac{e}{h} \int d\varepsilon \sum_{k=1,2,7,8}
  S^{(1)}_{j;k}(\varepsilon)^*   \nonumber \\ &\, \mbox{} \times
S^{(-1)}_{j;k}(\varepsilon+2\hbar \omega_{\rm M}) f_k(\varepsilon+\hbar \omega_{\rm M}). \label{eq:IacS2}
\end{align}

The steady-state precession frequency $\omega_{\rm M}$ is an unknown in this procedure and must be determined self-consistently using Eq.\ (\ref{eq:1}) and the steady-state condition
\begin{equation}
  \dot M_z(t) = 0.
  \label{eq:9}
\end{equation}
The net rate of change of $M_z$ is proportional to the net current reflected from the magnet ${\cal I}_{\rm r}$,\cite{meng2014,silvestrov2016}\textsuperscript{,}\footnote{We do not consider the case that the magnetization precession is driven by a source different from the applied voltage bias, such as an alternating magnetic field. In that case the magnet may function as a ``charge pump'' even in the absence of an applied bias voltage.\cite{qi2008} We also do neglect the effect of Gilbert damping, although both damping and external driving can be included relatively straightforwardly by adding additional terms to the right-hand side of Eq.\ (\ref{eq:MzI}).}
\begin{equation}
  \dot M_z(t) = \frac{{\cal I}_{\rm r}(t)}{e}. \label{eq:MzI}
\end{equation}
The Fourier transform ${\cal I}_{\rm r}(\omega)$ of the current reflected from the magnet is given by
\begin{align}
  {\cal I}_{\rm r}(\omega) =&\,
  \frac{e}{h}
  \int d\varepsilon
  \left[ \overline{\hat a^{\dagger}_{3+}(\varepsilon_+) \hat a_{3+}(\varepsilon_+ + \hbar \omega)} \right. \nonumber \\ &\, \ \ \ \ \left. \mbox{}
    - \overline{\hat a^{\dagger}_{6+}(\varepsilon_-) \hat a_{6+}(\varepsilon_- + \hbar \omega)} \right].
\end{align}
Electron paths contributing to ${\cal I}_{\rm r}(\omega)$ can differ by at most one reflection from the magnet, so that only Fourier compoments at $\omega = n \omega_{\rm M}$ with $n=-1,0,1$ contribute to ${\cal I}_{\rm r}(\omega)$,
\begin{equation}
  {\cal I}_{\rm r}(\omega) = \sum_{n=-1}^{1} I_{\rm r}(n \omega_{\rm M}) \delta(\omega - n \omega_{\rm M}).
\end{equation}
The Fourier components $I_{\rm r}(n \omega_{\rm M})$ depend on the precession frequency $\omega_{\rm M}$. They can be calculated using the scattering formalism outlined above.
The Fourier components $I_{\rm r}(n \omega_{\rm M})$ with $n = \pm 1$ give rise to small oscillations of $M_z$, which do not affect the precession frequency $\omega_{\rm M}$ for a macroscopic magnet. Keeping the zero-frequency component only, we find that the steady-state condition for $M_z$, from which the precession frequency $\omega_{\rm M}$ can be determined, reads
\begin{align}
  \label{eq:8}
  I_{\rm r}(0) = 0.
\end{align}

In the remaining sections we present explicit analytical and numerical results of this procedure for several representative choices of the scattering matrices $S^{{\rm C}1}$ and $S^{{\rm C}2}$ of the point contact regions.

\section{Right point contact ``open''}\label{sec:open}

\subsection{Calculation of DC and AC currents \label{ssec:opencurrents}}

\begin{figure}
	\centering
	\includegraphics[width=.21\textwidth]{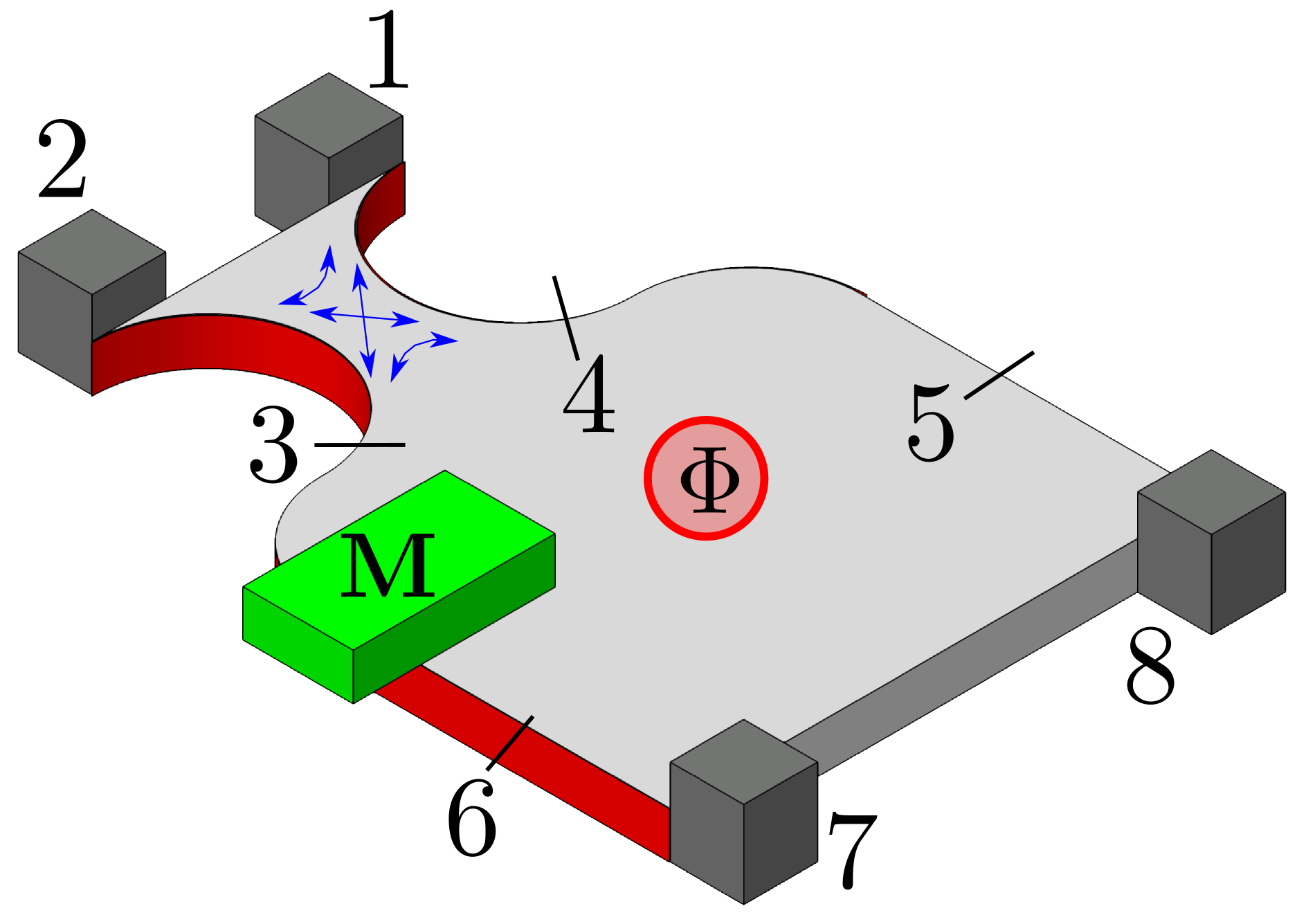}
	\includegraphics[width=.26\textwidth]{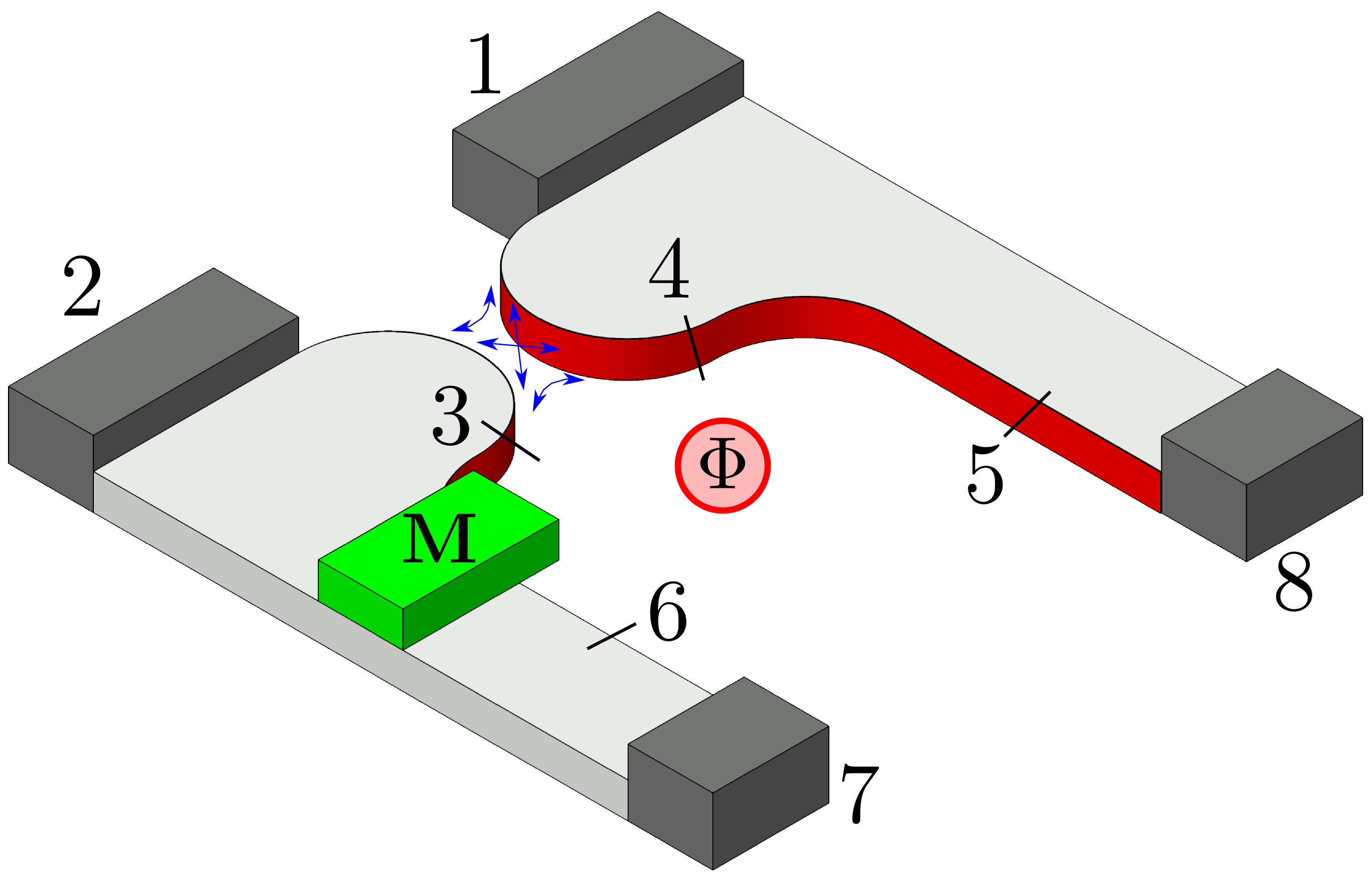}
	\caption{Schematic picture of the ``right point contact open'' geometry for a realization in which the interfering edge channels are on the same quantum spin Hall insulator (left) and on different quantum spin Hall insulators (right). The positions labeled $i=1,2,\ldots,8$ refer to the reference positions used for the calculations in the main text.}
	\label{fig:setup_open}
\end{figure}

As an analytically tractable geometry, we first consider a setup in which the right point-contact in Fig.\ \ref{fig:setup} is fully ``open,'' so that effectively there is a single point-contact region only (see Fig.\ \ref{fig:setup_open}). In this geometry, electrons incident from reservoirs ``1'', ``2'', and ``8'' can only reflect off the left end of the magnet, whereas electrons incident from reservoir ``7'' can only reflect off the right end of the magnet. This rules out interference processes with a total energy difference of $2 \hbar \omega_{\rm M}$, so that only harmonics $I_j(n \omega_{\rm M})$ with $n = 0$, $\pm 1$ need to be considered.

For the scattering matrix $S^{{\rm C}2}$ the condition that the right point contact is ``open'' implies $R_2 = T_2' = 0$, so that $\hat a_{5\pm}(\varepsilon) = \mp \hat a_{8\pm}(\varepsilon)$ and $\hat a_{6\pm}(\varepsilon) = \mp \hat a_{7\pm}(\varepsilon)$. We can express the remaining operators $\hat a_{j\pm}(\varepsilon)$ in terms of the operators $\hat a_{1+}(\varepsilon)$, $\hat a_{2+}(\varepsilon)$, $\hat a_{7+}(\varepsilon)$, and $\hat a_{8+}(\varepsilon)$ describing electrons coming in from the reservoirs by solving Eqs.\ (\ref{eq:4}) and (\ref{eq:6}) for this case. 

We first calculate the operators $\hat a_{3+}(\varepsilon)$ and $\hat a_{6+}(\varepsilon)$ describing electrons incident on the magnet, because these determine the precession frequency $\omega_{\rm M}$,
\begin{align}
   \hat a_{3+}(\varepsilon) =&\,
   i \hat a_{1+}(\varepsilon) \sqrt{T_1'}
   + \hat a_{2+}(\varepsilon) \sqrt{T_1}
   \nonumber \\ &\, \mbox{}
   - \hat a_{8+}(\varepsilon) \sqrt{R_1} 
   e^{i k(\varepsilon) L - i \phi}, \\
   \hat a_{6+}(\varepsilon) =&\, - \hat a_{7+}(\varepsilon).
\end{align}
The steady-state precession frequency $\omega_{\rm M}$ can be calculated from Eq.\ (\ref{eq:8}) upon setting $\dot M_z = 0$,
\begin{align}
   0 =&\,
   \int d\varepsilon |r_{\rm M}(\varepsilon)|^2
   \left[ f_{3+}(\varepsilon_+) - f_7(\varepsilon_-) \right],
   \label{eq:omegaMopen}
\end{align}
where we abbreviated
 \begin{equation}
   f_{3+}(\varepsilon) = R_1 f_8(\varepsilon) + T_1 f_2(\varepsilon) + T_1' f_1(\varepsilon)  .
   \label{eq:f3plus}
\end{equation}
The results for the operators $\hat a_{j-}(\varepsilon)$ in the leads ($j=1,2,7,8$) are then expressed in terms of the matrices of coefficients $S^{(n)}_{j;k}(\varepsilon)$ [see Eq.\ (\ref{eq:Sdef})], which read
\begin{widetext}
\begin{align}
  S^{(-)}(\varepsilon) =&\,
  r_{\rm M}(\varepsilon_-) \hat m_-
  \begin{pmatrix} 0 & 0 & 0 & 0 \\ 0 & 0 & 0 & 0 \\
  0 & 0 & -1 & 0 \\
  0 & 0 & 0 & 0 \end{pmatrix}, \nonumber \\
  S^{(0)}(\varepsilon) =&\,
  \begin{pmatrix}
  0 & \sqrt{R_1} & i t_{\rm M}'(\varepsilon_+) \sqrt{T_1'} &
  e^{i k L - i \phi} \sqrt{T_1} \\
  - \sqrt{R_1} & 0 & t_{\rm M}'(\varepsilon_+) \sqrt{T_1} &
  i e^{i k L - i \phi} \sqrt{T_1'} \\
  i t_{\rm M}(\varepsilon_-) \sqrt{T_1'} & t_{\rm M}(\varepsilon_-) \sqrt{T_1} & 0 & t_{\rm M}(\varepsilon_-) e^{i k L - i \phi} \sqrt{R_1} \\
  e^{i k L + i \phi} \sqrt{T_1} & i e^{i k L + i \phi} \sqrt{T_1'} & -t_{\rm M}'(\varepsilon_+)  e^{i k L + i \phi}  \sqrt{R_1} & 0
\end{pmatrix}, \\
  S^{(+)}(\varepsilon) =&\,
  r_{\rm M}(\varepsilon_+) \hat m_+
  \begin{pmatrix}
  T_1' & -i \sqrt{T_1 T_1'} & 0 & i e^{i k_+ L - i \phi} \sqrt{T_1' R_1} \\
  - i \sqrt{T_1 T_1'} & - T_1 & 0 & e^{i k_+ L - i \phi} \sqrt{T_1 R_1} \\
  0 & 0 & 0 & 0 \\
  - i e^{i k + i \phi} \sqrt{T_1' R_1} & - e^{i k L + i \phi} \sqrt{T_1 R_1} & 0 & e^{i (k+k_+) L} R_1 \end{pmatrix}, \nonumber
\end{align}
\end{widetext}
where we abbreviated $\varepsilon_{\pm} = \varepsilon \pm (1/2) \hbar \omega_{\rm M}$, $k = k(\varepsilon)$, and $k_\pm = k(\varepsilon \pm \hbar \omega_{\rm M})$.

For the currents in the four leads we then obtain the DC components from Eq.\ (\ref{eq:IdcS}): 
\begin{align}
\begin{split}
  I_1(0) =&\, \frac{e}{h} \int d\varepsilon
  \left[ f_1(\varepsilon)
  - R_1 f_{2}(\varepsilon)
  - T_1 f_{8}(\varepsilon)
  - T_1' f_7(\varepsilon) 
  \right], \\ 
  I_2(0) =&\, \frac{e}{h} \int d\varepsilon
  \left[ f_2(\varepsilon)
  - R_1 f_{1}(\varepsilon)
  - T_1 f_{7}(\varepsilon)
  - T_1' f_8(\varepsilon) 
  \right], \\
  I_7(0) =&\, \frac{e}{h} \int d\varepsilon
  \left[ f_7(\varepsilon) 
  - f_{3+}(\varepsilon)
  \right], \\
  I_8(0) =&\, \frac{e}{h} \int d\varepsilon
  \left[ f_8(\varepsilon)
  - R_1 f_{7}(\varepsilon)
  - T_1 f_{1}(\varepsilon)
  - T_1' f_2(\varepsilon) 
  \right], 
  \label{eq:Idcopen}
\end{split}
\end{align}
where $f_{3+}(\varepsilon)$ is given in Eq.\ (\ref{eq:f3plus}). Since the DC current components do not depend on the reflection amplitude $r_{\rm M}$ of the magnet---which is consistent with the observation that the magnet does not reflect current in the steady-state regime\cite{meng2014,silvestrov2016}---the integrations over energy can be carried out explicitly and one finds the simple result
\begin{align}
\begin{split}
  I_1(0) =&\, \frac{e^2}{h} (V_1 - R_1 V_2 - T_1 V_8 - T_1' V_7), \\
  I_2(0) =&\, \frac{e^2}{h} (V_2 - R_1 V_1 - T_1 V_7 - T_1' V_8), \\
  I_7(0) =&\, \frac{e^2}{h} (V_7 - R_1 V_8 - T_1 V_2 - T_1' V_1), \\
  I_8(0) =&\, \frac{e^2}{h} (V_8 - R_1 V_7 - T_1 V_1 - T_1' V_2).
\end{split}
\end{align}
Interference between transmission paths that reflect from the magnet and paths that do not reflect from the magnet gives an AC contribution to the current at frequency $\omega = \pm \omega_{\rm M}$. For the AC components at frequency $\omega = \omega_{\rm M}$ we find 
\begin{align} \label{eq:IACopen}
  I_1(\omega_{\rm M}) =&\, - i \hat m_- \frac{e}{h} \sqrt{T_1 T_1' R_1} \int d\varepsilon
  r_{\rm M}(\varepsilon)^* 
  \nonumber \\ &\, \mbox{} \times
  \left[
  f_2(\varepsilon_+) - f_8(\varepsilon_+) \right], \nonumber \\
  I_2(\omega_{\rm M}) =&\, - i \hat m_- \frac{e}{h} \sqrt{T_1 T_1' R_1} \int d\varepsilon
  r_{\rm M}(\varepsilon)^*
  \nonumber \\ &\, \mbox{} \times
  \left[
  f_8(\varepsilon_+) - f_1(\varepsilon_+) \right], \nonumber \\
  I_7(\omega_{\rm M}) =&\, 0, \\
  I_8(\omega_{\rm M}) =&\, - i \hat m_- \frac{e}{h} \sqrt{T_1 T_1' R_1} \int d\varepsilon
  r_{\rm M}(\varepsilon)^*  
  \nonumber \\ &\, \mbox{} \times
  e^{i k(\varepsilon_+) L - i k(\varepsilon_-) L} 
  \left[ f_1(\varepsilon_+) - f_2(\varepsilon_+) \right]. \nonumber
\end{align}
[The components at $\omega = -\omega_{\rm M}$ are obtained by complex conjugation, $I_j(-\omega_{\rm M}) = I_j(\omega_{\rm M})^*$.] One verifies that the DC currents sum to zero. Current conservation also applies to the AC current components, if one corrects $I_8(\omega_{\rm M})$ for the time delay accumulated during the propagation along the upper interferometer arm.

\begin{figure}
\centering
\includegraphics[width=0.4\textwidth]{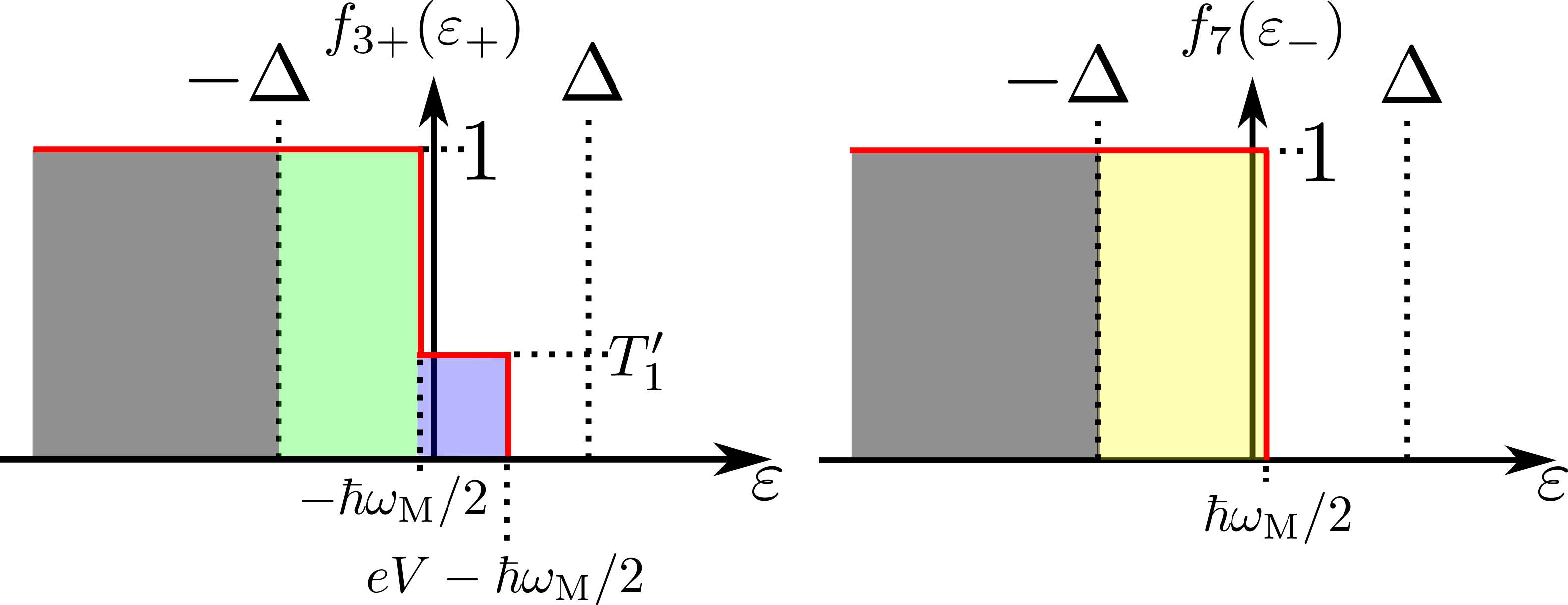}
\includegraphics[width=0.4\textwidth]{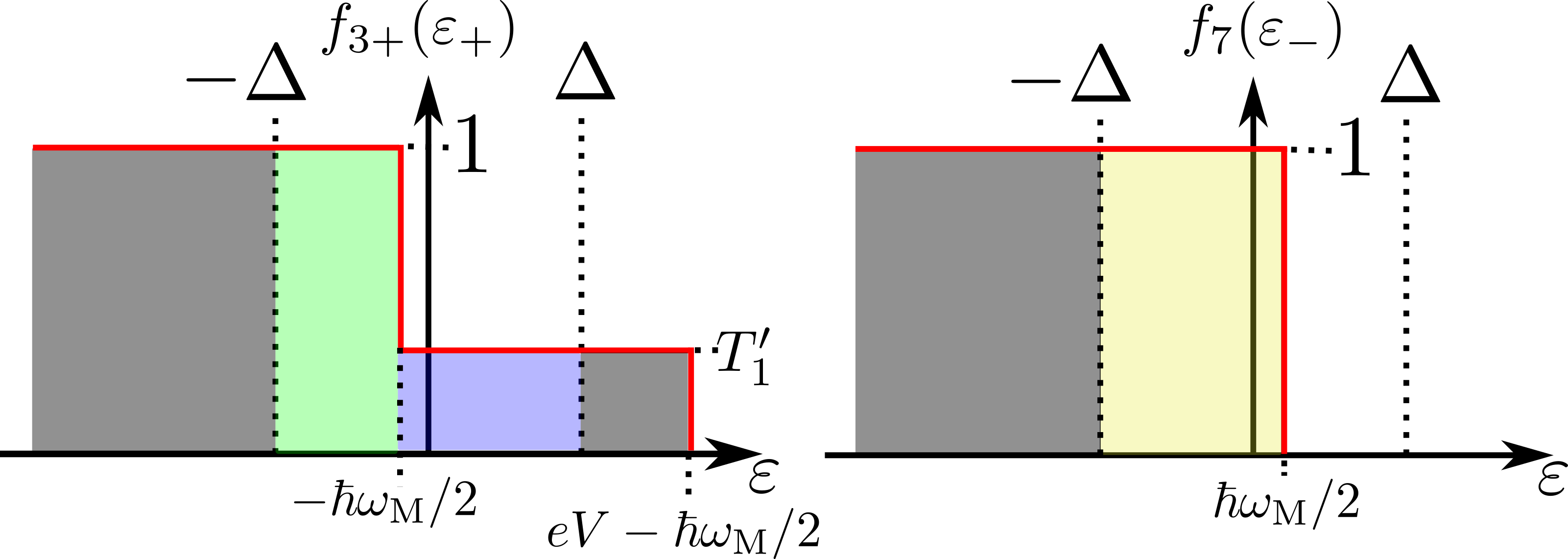}
\caption{\label{fig:distributions} Distribution functions $f_{3+}(\varepsilon_+)$ and $f_{7}(\varepsilon_-)$ for bias voltage $e V < 2\ \Delta'/(2 - T_1')$ (top panel) and $e V > 2 \Delta'/(2 - T_1')$ (bottom panel). The bias voltage $V$ is applied to lead ``1'' only; the (electro)chemical potentials of leads ``2,'' ``7,'' and ``8'' are held constant at the value $\mu = 0$.}
\end{figure}

To make the results for the AC current component more explicit, we now consider a simple model for the reflection from the magnetic insulator,
\begin{equation}
  r_{\rm M}(\varepsilon) = e^{2 i k(\varepsilon) L_3}\theta(\Delta - |\varepsilon|), \label{eq:rMopen}
\end{equation}
where $\Delta$ is the magnitude of the magnet-induced exchange gap in the helical edge, $L_3$ is the length of the interferometer arm between the left point contact and the magnet, and the Heaviside function $\theta(x) = 1$ if $x > 0$ and $0$ otherwise. We further set the temperatures $T_i$ to zero, choose $V_1 = V > 0$, $V_2 = V_7 = V_8 = 0$, and approximate $k(\varepsilon) = k_{\rm F} + \varepsilon/\hbar v_{\rm F}$. Solving the precession frequency $\omega_{\rm M}$ from Eq.\ (\ref{eq:omegaMopen}) then gives 
\begin{equation}
  \hbar \omega_{\rm M} = T_1'\, \min \left(e V, \frac{2 \Delta}{2 - T_1'} \right). 
	\label{eq:omegaMopenapprox}
\end{equation}
Schematic pictures of the distribution functions $f_{3+}(\varepsilon_+)$ and $f_7(\varepsilon_-)$ for $eV < 2 \Delta/(2- T_1')$ and $eV > 2 \Delta/(2-T_1')$ are shown in Fig.\ \ref{fig:distributions}.

In the limit $T_1' \rightarrow 1$ the right point contact ``open'' geometry studied in this section corresponds to the two-terminal setup studied in Ref. \onlinecite{silvestrov2016}. However, in that work the precession frequency is found to be $\hbar \omega_{\rm M} = e V$ and does not saturate for $e V > 2 \Delta$, as it does here [see Eq.\ (\ref{eq:omegaMopenapprox})]. This difference can be understood as follows: In the two-terminal geometry of Ref. \onlinecite{silvestrov2016}, when a stationary state of the magnet is reached, the current reflected by the magnet that adds angular momentum to the magnet is balanced by the current that removes angular momentum from the magnet. Both reflected currents are proportional to $|r_{\rm M}(\varepsilon)|^2$. Since it is assumed that the reflection amplitude $r_{\rm M}(\varepsilon)$ is nonzero for all energies, $r_{\rm M}(\varepsilon)$ drops out from the stationarity condition, giving the result $\hbar \omega_{\rm M} = e V$. In the four-terminal geometry studied in the present work, the magnet has additional channels available, through which it can ``lose'' angular momentum. These loss channels dominate over the gain-channel in our simple model, Eq. (\ref{eq:rMopen}), where $|r_{\rm M}(\varepsilon)|^2\rightarrow0$ for $\varepsilon > \Delta$. Hence, $\omega_{\rm M}$ saturates for sufficiently large $e V$. If we were to take the limit $T_1' \rightarrow 1$ in the beginning of the calculation, thereby {\em a priori} reducing the four-terminal geometry to the two-terminal geometry of Ref. \onlinecite{silvestrov2016}, before making any additional assumptions about the reflection amplitude $r_{\rm M}(\varepsilon)$, we would find the same result $\hbar \omega_{\rm M} = e V$ as in Ref. \onlinecite{silvestrov2016}.

For the AC current components we then find $I_1(\omega_{\rm M}) = I_7(\omega_{\rm M}) = 0$ and
\begin{align}
\label{eq:ACcurrentopen}
  I_2(\omega_{\rm M}) =&\,   -I_8(\omega_{\rm M}) e^{-i \omega_{\rm M} L/v_{\rm F}} \nonumber \\ =&\,
  \frac{e i v_{\rm F}}{2 \pi L_3} \hat m_- \sqrt{T_1 T_1' R_1} 
  \sin \frac{\omega_{\rm M} L_3}{v_{\rm F} T_1'}
  \nonumber \\ &\, \mbox{} \times
e^{-i L_3 [2 k_{\rm F}  + \omega_{\rm M}(R_1+T_1)/T_1' v_{\rm F}]}.
\end{align}
For a small applied bias and/or for a small induced gap $\Delta$, one may approximate $\omega_{\rm M} L_3/v_{\rm F} T_1'$, $\omega_{\rm M} L/v_{\rm F} \ll 1$ for typical device sizes, so that one has
\begin{align}
  I_2(\omega_{\rm M}) \approx&\, -I_8(\omega_{\rm M}) \nonumber \\
  \approx&\, \frac{e i \omega_{\rm M}}{2 \pi T_1'} \hat m_- \sqrt{T_1 T_1' R_1} e^{-2 i k_{\rm F} L_3}.
  \label{eq:I2ACexample}
\end{align}

For a discussion of the physical origin of the alternating current contribution in this interferometer, we focus on $I_2$ and note that the transmission amplitude between contacts ``1'' and ``2'' is the sum of two interfering contributions: A direct contribution with amplitude $\propto \sqrt{R_1}$ and a contribution $\propto \hat m_- \sqrt{T_1' T_1}$ with electrons reflected off the precessing magnet. Because the reflection phase depends on the magnetization direction (via the expectation value of ${\hat m}_{-}$), electrons that reflect from the magnet pick up a time-dependent phase factor $\propto e^{- i \omega_{\rm M} t}$. As a result, the interference contribution to the transmission probability between the reservoirs ``1'' (at which the voltage bias is applied) and ``2'' (at which the current is measured) oscillates with frequency $\omega_{\rm M}$. At a constant bias voltage, this periodic modulation of the (Landauer-B\"uttiker) transmission probability between reservoirs leads to an alternating contribution to the current $I_2$ at frequency $\omega_{\rm M}$.

That the alternating current component is a time-dependent interference-related modulation of the transmission between reservoirs and not the consequence of a periodic spin current pumped by a precessing magnetization [\onlinecite{tserkovnyak2002}] and then converted into a charge current by spin-dependent scattering, can be seen from the presence of the additional phase factor $e^{-2 i k_F L_3}$ and the corresponding suppression of the alternating current component if $eV \gtrsim \hbar L_3/v_{\rm F}$ [see Eq.\ (\ref{eq:ACcurrentopen})]. Such phase factors are absent in the theory of spin pumping and related phenomena [\onlinecite{tserkovnyak2005}]. It can also be seen by comparing the magnitudes of the alternating current component $I_2(\omega_{\rm M})$ and the direct current $I_2(0)$. Unlike the AC contribution, the direct current may be interpreted in terms of pumping by the precessing magnetization.\cite{meng2014} For the two-terminal geometry of Refs.\ \onlinecite{meng2014,silvestrov2016}, the precession frequency of the magnet equals the applied bias, $\hbar \omega_{\rm M} = e V$, so that in the steady state there is exactly one electron transferred in a period of the precessing magnetization. (This places the throughput of this device in the same category as the electron current pumped through a Coulomb-blockaded quantum dot.\cite{thouless1983,kouwenhoven1991,pothier1992}) The same universal estimate, one electron per period, also holds for the DC currents in the interferometer devices considered in our paper, when applied to the contribution of the DC current $I_3(0) = I_6(0) = I_7(0)$ transmitted through the magnet for bias voltage $|eV| < \Delta$ below the magnet-induced gap. The AC current component $I_2(\omega_{\rm M})$, however, may be much larger than this, depending on the values of the transmission coefficients $T_1$, $T_1'$, and $R_1$: From Eq.\ (\ref{eq:I2ACexample}) one sees that $I_2(\omega_{\rm M})$ may be as large as a charge $\sim e \sqrt{R_1 T_1/T_1'}$ per period of the precessing magnetization. This charge may be much larger than $e$ in the limit $T_1' \ll 1$, ruling out pumping as an origin of $I_2(\omega_{\rm M})$. [The AC current $I_2(\omega_{\rm M})$ is always
smaller than the total DC current $I_1(0)$ injected in lead ``1''.] We note that the limit $T_1' \ll 1$ is a physically relevant limit, since, as discussed in Sec.\ \ref{setup}, for scattering between helical edges with equal spin polarization $T_1'$ is related to spin-flip processes. Such processes originate from the Rashba spin-orbit interaction, sensitive to the inversion-breaking electric field in the contact region. Since such a field is intrinsically weak compared to the crystal fields, one may expect the spin-flip processes to be suppressed, hence $T_1' \ll 1$. On the other hand, $T_1'$ may be effectively tuned by external electric fields.

\subsection{``Classical'' vs.\ ``quantum'' magnet}\label{ssec:cl-vs-qu}

The expressions for the AC contributions $I_2(\omega_{\rm M})$ given in Eq.\ (\ref{eq:ACcurrentopen}) are proportional to the lowering operator $\hat{m}_-$. The proper interpretation of these equations requires a brief discussion how expressions for (expectation values of) current components $I_j(n \omega_{\rm M})$ that contain the operators $\hat{m}_-$ and $\hat{m}_+$ should be understood. Hereto, we distinguish the ``classical'' case that $\vM$ is (effectively) a classical vector with a well-defined direction and the ``quantum'' case that $\vM$ must be considered an operator and the magnet is in a state with well-defined $M_z$, whereas the components $M_x$ and $M_y$ are maximally uncertain.

If the magnet is in a ``classical'' state $|C\rangle$ describing a macroscopic magnetization, then the magnetization has a well-defined direction at any time $t$, parametrized by the polar angle $\theta$ with the $z$ axis and the azimuthal angle $\varphi$. In that case, although $|C\rangle$ is not an eigenstate of any component of the magnetization operator $\vec M$, all three components of $\vec M$ have a finite expectation value. In particular, the expectation value $\langle C|\hat{m}_- |C\rangle \equiv \langle \hat{m}_- \rangle_{C}$ of the projection of $\vM$ on the $xy$ plane takes the nonzero value
\begin{equation}
  \langle \hat{m}_- \rangle_{C} = e^{i \varphi}.
  \label{eq:mC}
\end{equation}
[Note that the ladder operators $\hat m_{\pm}$ are defined as $\hat{m}_{\pm} = (m_x \pm i m_y)/|m_x \pm i m_y|$, which is why the expectation value (\ref{eq:mC})  does not depend on the polar angle $\theta$.]

If the magnet is in a ``quantum'' state $|Q\rangle$ with a sharply-defined quantized value of $M_z$, the azimuthal angle $\varphi$ of the magnetization is unknown. In this case one must average over all phase angles $\varphi$,  so that the expectation value $\langle \hat{m}_- \rangle_Q = 0$. Then, the quantity that should be considered to characterize the AC currents is not the expectation value $\langle I_j(\omega) \rangle_Q$, but the current correlations
\begin{equation}
  \langle I(\omega_{\rm M}) I(-\omega_{\rm M}) \rangle_Q = \langle I(\omega_{\rm M}) I^{\dagger}(\omega_{\rm M}) \rangle_Q.
\end{equation}
The current correlations are proportional to $\langle \hat{m}_- \hat{m}_+\rangle_Q = 1$. For example, from Eq.\ (\ref{eq:ACcurrentopen}) one then obtains
\begin{align}
  \langle I_2(\omega_{\rm M}) I_2^{\dagger}(\omega_{\rm M}) \rangle_Q =&\, \frac{e^2 v_{\rm F}^2}{4 \pi^2 L_3^2} T_1 T_1' R_1 
\sin^2 \frac{\omega_{\rm M} L_3}{v_{\rm F} T_1'} .
\end{align}
In the same way, the other results for AC currents in Sec.\ \ref{sec:open} and in the following sections can be interpreted in terms of a current correlator. 
Note that the exchange of energy between the magnet and the helical edge state electrons upon reflection from the magnet is given by Eqs. (\ref{eq:6})-(\ref{eq:9}), which are independent of the state of the magnet (``classical'' vs ``quantum''). Therefore the analysis of the scattering problem works in both cases discussed in this section.

\section{Right point contact ``closed''}\label{sec:closed}

\begin{figure}
	\centering
	\includegraphics[width=.21\textwidth]{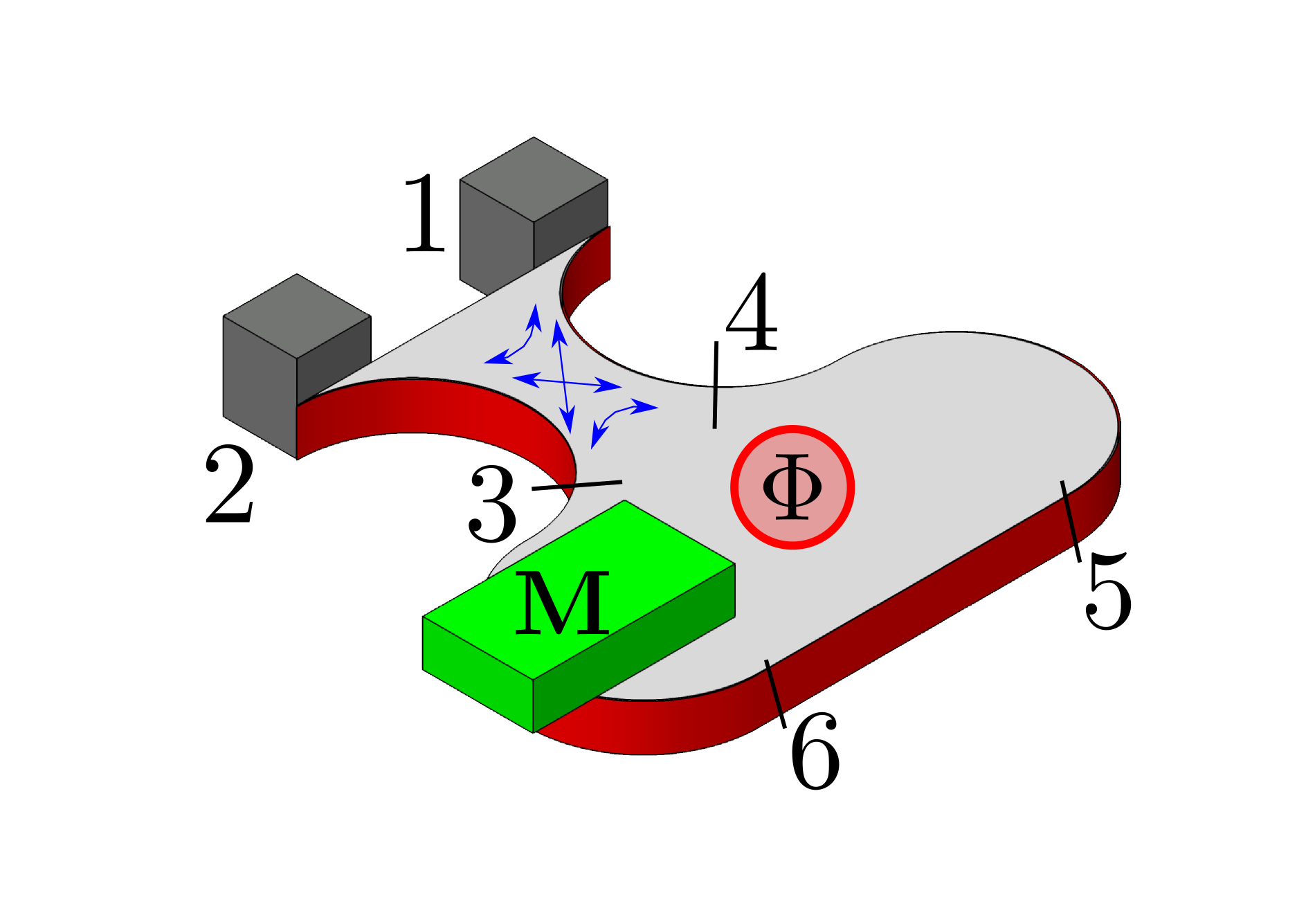}
	\includegraphics[width=.26\textwidth]{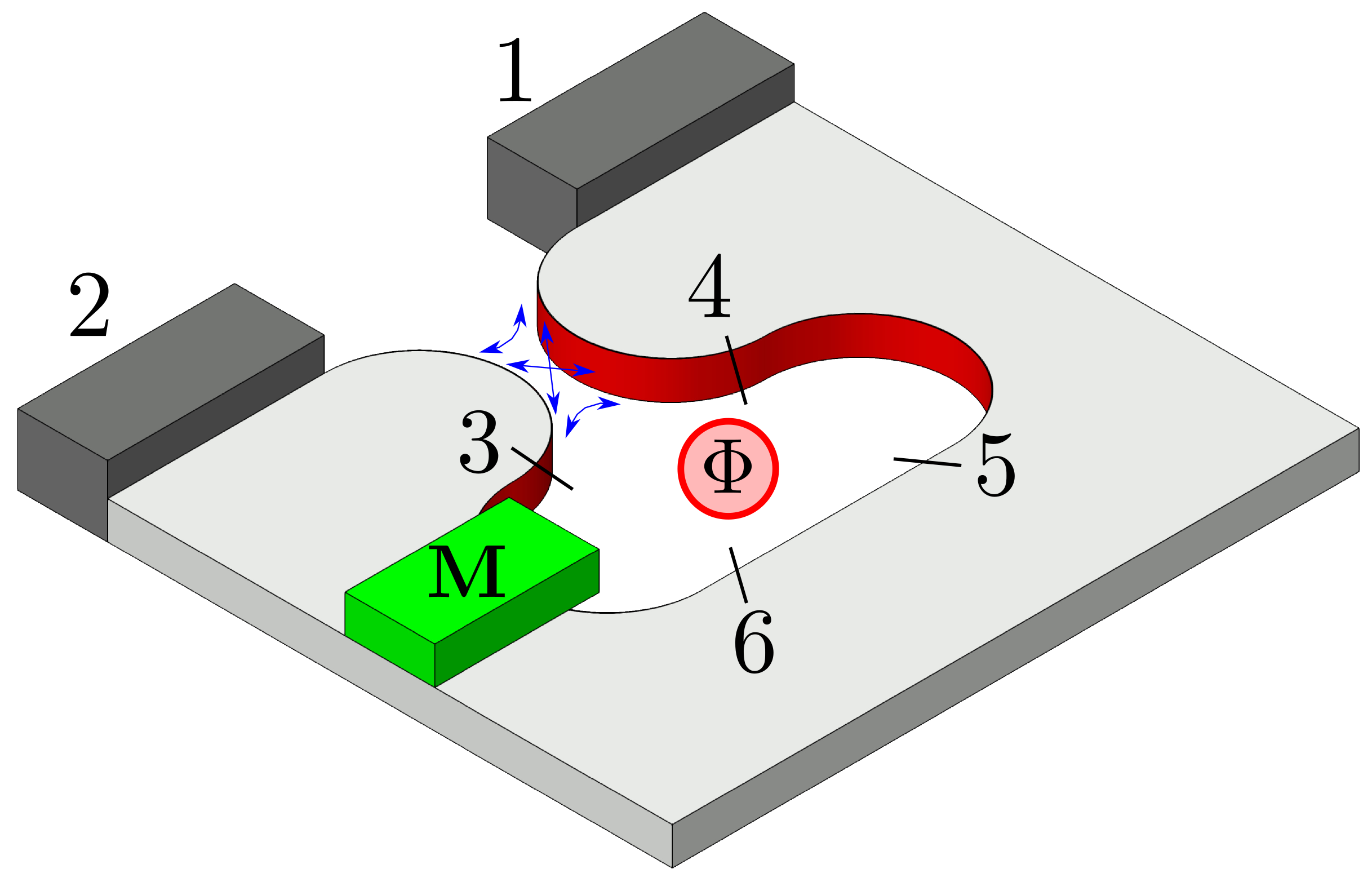}
	\caption{Schematic picture of the ``right point contact closed'' geometry for a realization in which the interfering edge channels are on the same quantum spin Hall insulator (left) and on different quantum spin Hall insulators (right). The positions labeled $i=1,2,\ldots,8$ refer to the reference positions used for the calculations in the main text.}
	\label{fig:setup_closed}
\end{figure}

As a second example we consider a setup in which the point contact to the right of the magnet is fully ``closed'' (see Fig.\ \ref{fig:setup_closed}). Although this is effectively a two-terminal geometry (in contrast to the four-terminal geometry of the previous section), it has a richer phenomenology, since it allows for AC current components at frequencies $\omega_{\rm M}$ as well as $2 \omega_{\rm M}$. The current component at frequency $2 \omega_{\rm M}$ comes from interference of electrons reflecting off the left end of the magnet (where energy is decreased by $\hbar \omega_{\rm M}$ upon reflection) and the right end of the magnet (where an energy quantum $\hbar \omega_{\rm M}$ is absorbed upon reflection), so that the net energy difference in the interference process is $2 \omega_{\rm M}$.

For the scattering matrix $S^{{\rm C}2}$ the condition that the right point contact is ``closed'' translates to $R_2 = 1$, $T_2=T_2'=0$, so that $\hat{a}_{5\pm}(\varepsilon)=\pm \hat{a}_{6\mp}(\varepsilon)$. As we will show below, in this setup the current has AC components at frequencies $\omega_{\rm M}$ and $2 \omega_{\rm M}$. To keep the expressions simple, we will assume that the left point contact is close to being ``open,'' $R_1$, $T_1' \ll 1$, and give final expressions to lowest nontrivial order in $R_1$, $T_1'$. The leads labeled ``7'' and ``8'' are disconnected from the magnet and will not be considered here. 

By solving Eqs.\ (\ref{eq:4}) and (\ref{eq:6}), the operators $\hat{a}_{3+}(\varepsilon)$ and $\hat{a}_{6+}(\varepsilon)$ for electrons incident on the magnet can be expressed in terms of the operators $\hat{a}_{1+}(\varepsilon)$ and $\hat{a}_{2+}(\varepsilon)$ for electrons incoming from the reservoirs on the left side. To first order in $\sqrt{R_1}$, $\sqrt{T_1'}$ we find
\begin{widetext}
\begin{align}
\hat{a}_{3+}(\varepsilon_+) =&\,
  -  \hat{a}_{1+}(\varepsilon_-)
  e^{ik(\varepsilon_-)L+ik(\varepsilon_+)L} \hat{m}_- r_{\rm M}'(\varepsilon) \sqrt{R_1} + 
  \hat a_{2+}(\varepsilon_+) \left[ 1 + e^{-i\phi+ i k(\varepsilon_+)L} t_{\rm M}(\varepsilon) \sqrt{R_1} \right]
  + i \hat a_{1+}(\varepsilon_+) \sqrt{T_1'}
, \\
  \hat{a}_{6+}(\varepsilon_-) =&\,
  e^{i\phi + i k(\varepsilon_-)L} \left\{ 
  i \hat a_{2+}(\varepsilon_-) \sqrt{T_1'} 
  - \hat a_{1+}(\varepsilon_-) \left[1 +  e^{i \phi + i k(\varepsilon_-) L} t_{\rm M}'(\varepsilon) \sqrt{R_1} \right]
  - \hat a_{2+}(\varepsilon_+) \hat{m}_- r_{\rm M}(\varepsilon) \sqrt{R} \right\}.
\end{align}
The results for the operators $\hat a_{j-}(\varepsilon)$ for the outgoing modes in the leads $j=1,2$ are again expressed in terms of the matrices of coefficients $S^{(n)}_{j;k}(\varepsilon)$, which are effectively $2\rm x 2$-matrices for the geometry we consider here. These matrices read
\begin{align}
S^{(-)}(\varepsilon) =&\,
	r_{\rm M}'(\varepsilon_-) \hat m_- e^{i (k + k_-) L}	
	\begin{pmatrix} 
	  1 + \sqrt{R_1}
          [t_{\rm M}(\varepsilon_-) e^{-i (\phi - kL)} + t_{\rm M}'(\varepsilon_-) e^{i (\phi + k_- L)}] && i \sqrt{T_1'} \\
	i \sqrt{T_1'} && - T_1'   	
	\end{pmatrix}, \nonumber \\
  S^{(0)}(\varepsilon) =&\,
	e^{i k L}	
	\begin{pmatrix}
  	i \sqrt{T_1'} 
  [e^{i \phi} t_{\rm M}'(\varepsilon_+) - e^{-i \phi} t_{\rm M}(\varepsilon_-)] && -e^{-i \phi} t_{\rm M}(\varepsilon_-) \\
  	e^{i \phi} t_{\rm M}'(\varepsilon_+) && i \sqrt{T_1'} [e^{i \phi} t_{\rm M}'(\varepsilon_+) - e^{-i \phi} t_{\rm M}(\varepsilon_-)]  \\
	\end{pmatrix}
  \nonumber \\ &\, \mbox{}
 + \sqrt{R_1}  
  \begin{pmatrix}
  	\sqrt{T_1'} \sum_{\pm} [\pm e^{i (k+k_{\pm}) L} r_{\rm M}(\varepsilon_{\pm}) r_{\rm M}'(\varepsilon_{\pm}) ]
& i [e^{i (k+k_-) L} r_{\rm M}(\varepsilon_-) r_{\rm M}'(\varepsilon_-) - 1] \\
	 i [1 - e^{i (k+k_+) L} r_{\rm M}(\varepsilon_+) r_{\rm M}'(\varepsilon_+)] & \sqrt{T_1'} \sum_{\pm} [\pm e^{i (k+k_{\pm}) L} r_{\rm M}(\varepsilon_{\pm}) r_{\rm M}'(\varepsilon_{\pm}) ] \\
	\end{pmatrix},   \nonumber \\
  S^{(+)}(\varepsilon) =&\,
	r_{\rm M}(\varepsilon_+) \hat m_+	
	\begin{pmatrix}
	T_1' && -i \sqrt{T_1'} \\
  	-i \sqrt{T_1'} && -1 - \sqrt{R_1} [t_{\rm M}(\varepsilon_+) e^{-i (\phi - k_+L)} + t_{\rm M}'(\varepsilon_+) e^{i (\phi + k L)}]
	\end{pmatrix}, 
\end{align}
where we again abbreviated $\varepsilon_{\pm} = \varepsilon \pm (1/2) \hbar \omega_{\rm M}$, $k=k(\varepsilon)$, and $k_\pm = k(\varepsilon \pm \hbar \omega_{\rm M})$. We have kept contributions beyond the lowest order in $\sqrt{R_1}$ and $\sqrt{T_1'}$ as far as these are important for a consistent expansion of the currents for small $\sqrt{R_1}$ and $\sqrt{T_1'}$ and for small bias voltage.

We proceed to calculate the steady-state precession frequency $\omega_{\rm M}$ and the currents $I_1$ and $I_2$. 
The steady-state precession frequency $\omega_{\rm M}$ is calculated from Eq.\ (\ref{eq:8}) upon setting $\dot M_z = 0$. In the limit $T_1'$, $R_1 \ll 1$ Eq.\ (\ref{eq:8}) gives
\begin{align}
  0 =&\,
  \int d\varepsilon |r_{\rm M}(\varepsilon)|^2 [ f_2(\varepsilon_+)-f_1(\varepsilon_-) ],
  \label{eq:omegaMclosed}
\end{align}
from which $\omega_{\rm M}$ can be determined. (Explicit results for a simple model will be given below.) The DC current components $I_1(0)$ and $I_2(0)$ now acquire a weak dependence on the reflection amplitude $r_{\rm M}$ of the magnet,
\begin{align}
      \label{eq:IDCclosed}
  I_1(0) =&\, - I_2(0) \nonumber \\
  =&\, \frac{e^2}{h} (V_1-V_2) 
  + \frac{e}{h} \sqrt{R_1} \int d\varepsilon
  |r_{\rm M}(\varepsilon_-)|^2 [f_2(\varepsilon)-f_1(\varepsilon-\hbar \omega_{\rm M})]
  \nonumber \\ &\, \mbox{} \times
  \left\{ t_{\rm M}'(\varepsilon_-) e^{i(\phi+k_- L)} + t_{\rm M}'^*(\varepsilon_-) e^{-i(\phi+k_- L)}
    + t_{\rm M}(\varepsilon_-) e^{-i(\phi-k L)} + t_{\rm M}^*(\varepsilon_-) e^{i(\phi-k L)} \right\},
\end{align}
where we kept only those correction terms that depend on the Aharonov-Bohm phase $\phi$. Note that $\omega_{\rm M} = 0$ if $V_1 = V_2$, so that$I_j(0) = 0$, $j=1,2$ in the absence of an applied bias. The AC components at frequency $\omega = \omega_{\rm M}$ are
\begin{align}
\begin{split}
  I_1(\omega_{\rm M}) =&\, i \hat{m}_- \sqrt{T_1'} \frac{e}{h} \int d\varepsilon\, \left\{ 
  r_{\rm M}'(\varepsilon_-) 
  t_{\rm M}'^*(\varepsilon_-) e^{-i \phi} e^{i k L} \left[ f_1(\varepsilon - \hbar \omega_{\rm M}) - f_2 (\varepsilon) \right] - r_{\rm M}'(\varepsilon_{+}) e^{i \phi} e^{i k_+ L} t_{\rm M}^*(\varepsilon_{-}) 
  \left[ f_1(\varepsilon)-f_2(\varepsilon)) \right]  \right\} \\
  &\,+i \hat{m}_- \sqrt{T_1' R_1} \frac{e}{h}\int d\varepsilon\,
  \left\{
  -e^{i(k_+-k_-)L}r_{\rm M}'(\varepsilon_+)r_{\rm M}'^*(\varepsilon_-)r_{\rm M}^*(\varepsilon_-) \left[ f_1(\varepsilon) - f_2(\varepsilon) \right] \right.
\\  &\, \ \ \ \ \mbox{} +
  r_{\rm M}^*(\varepsilon_-) \left[ f_1(\varepsilon-\hbar \omega_{\rm M}) - f_2(\varepsilon) \right] 
  - \left.
  e^{i(k_-+k)L}r_{\rm M}'(\varepsilon_-) \left[f_2(\varepsilon-\hbar \omega_{\rm M}) -  f_2(\varepsilon) \right]  \right\},
\\
  I_2(\omega_{\rm M}) =&\, i \hat{m}_- \sqrt{T_1'} \frac{e}{h} \int d\varepsilon\,
   \left\{ r_{\rm M}^*(\varepsilon_-) 
  t_{\rm M}(\varepsilon_-) e^{-i \phi} e^{i k L}  \left[ f_1(\varepsilon-\hbar \omega_{\rm M}) - f_2(\varepsilon) \right] - r_{\rm M}^*(\varepsilon_{-}) e^{i\phi} e^{i k L} t_{\rm M}'(\varepsilon_{+}) \left[ f_1(\varepsilon) -f_2(\varepsilon)  \right]  \right\}\\
  &\,+i \hat{m}_- \sqrt{T_1' R_1} \frac{e}{h}\int d\varepsilon\,
  \left\{
  -e^{i(k+k_+)L} r_{\rm M}(\varepsilon_+)r_{\rm M}'(\varepsilon_+)r_{\rm M}^*(\varepsilon_-) \left[ f_1(\varepsilon) - f_2(\varepsilon) \right] \right.
\\ 
&\, \ \ \ \ \mbox{} +
  e^{i(k_-+k)L} r_{\rm M}'(\varepsilon_-) \left[ f_1(\varepsilon-\hbar \omega_{\rm M}) - f_2(\varepsilon) \right]
  - \left. \vphantom{e^{i(k+k_+)L}}
  r_{\rm M}^*(\varepsilon_-) \left[ f_1(\varepsilon-\hbar \omega_{\rm M}) - f_1(\varepsilon) \right]  \right\},
  \label{eq:IomegaMclosed}
  \end{split}
\end{align}
\begin{figure}[h!]
	\centering
	\includegraphics[width=0.45\textwidth]{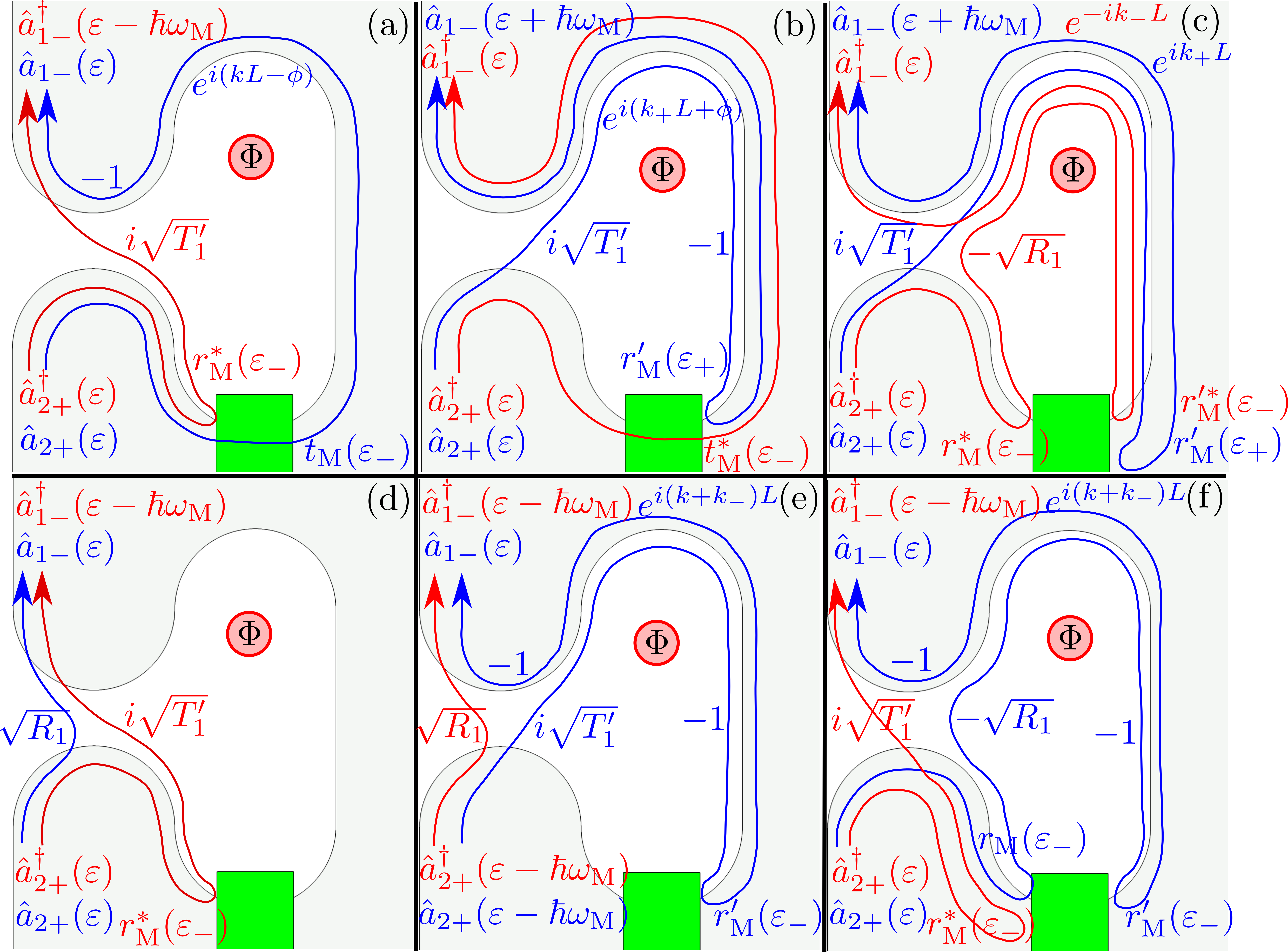}
	\caption{\label{fig:closedAC} Contributions to the AC current amplitude $I_{1}(\omega_{\rm M})$ in lead ``1'' that depend on the distribution function $f_2(\varepsilon)$ in lead ``2.'' The panels (a)--(f) are in the same order as the terms of Eq. (\ref{eq:IomegaMclosed}), but when attempting to match the expressions from the panels to Eq. (\ref{eq:IomegaMclosed}) care must be taken as explained in the following. An overall minus sign that comes from the definition of the current $I_1(\omega)$ in Eq. (\ref{eq:Iomega}) is not included. In panel (a) unitarity of $S^{({\rm M})}$ must be used to make the replacement $-r_{\rm M}^*(\varepsilon_-)t_{\rm M}(\varepsilon_-) = r_{\rm M}'(\varepsilon_-)t_{\rm M}'^*(\varepsilon_-)$ to arrive at the corresponding term in Eq.\ (\ref{eq:IomegaMclosed}). In comparison to Eq.\ (\ref{eq:IomegaMclosed}), the contribution from panel (f) contains an extra factor of $|r_{\rm M}(\varepsilon_-)|^2$. This factor has been dropped in Eq.\ (\ref{eq:IomegaMclosed}), as $\sqrt{T_1' R_1}r_{\rm M}'(\varepsilon_-)|r_{\rm M}(\varepsilon_-)|^2 \approx \sqrt{T_1' R_1}r_{\rm M}'(\varepsilon_-)$ to the level of approximation made in Eq.\ (\ref{eq:IomegaMclosed}). In panels (c) and (f) the property $\hat{m}_- \hat{m}_+ = \mathbbm{1}$ is used, as the red or blue paths reflect off both sides of the magnet. The factors of $-1$, $\sqrt{R_1}$, and $i\sqrt{T_1'}$ printed along the paths come from the scattering matrix elements of  $S^{({\rm C}1)}$ and $S^{({\rm C}2)}$ related to the paths that are shown in the limits considered.}
\end{figure}
where we have also included a subleading contribution in the expansion for $\sqrt{T_1'}$, $\sqrt{R_1}$, which we consider, because the leading contribution vanishes for small bias voltage if the Fermi energy is in the magnet-induced spectral gap. 
As an illustration, the six configurations of transmission paths that give the $f_2(\varepsilon)$-dependent contributions to $I_{1}(\omega_{\rm M})$ are shown in Fig.\ \ref{fig:closedAC}. The AC components at frequency $\omega = 2\omega_{\rm M}$ are
\end{widetext}
\begin{align}
  I_1(2\omega_{\rm M}) =&\, 
  - I_2(2 \omega_{\rm M}) \nonumber \\ =&\,
  i \hat{m}_-^2 T_1' \frac{e}{h} \int d\varepsilon
  e^{i (k+k_+) L} 
  r_{\rm M}'(\varepsilon_{+}) \nonumber \\ &\, \ \ \mbox{} \times
  r_{\rm M}^*(\varepsilon_-) \left[ f_2(\varepsilon) - f_1(\varepsilon)  \right].
  \label{eq:I2omegaMclosed}
\end{align}
These AC current components at frequency $2 \omega_{\rm M}$ can be understood just as in Fig.\ \ref{fig:closedAC}, the
difference being that the red and blue paths must now be reflected from opposite sides of the magnet so that their total energy difference is $2 \hbar \omega_{\rm M}$ and not $\hbar \omega_{\rm M}$, as it is for the interference partners shown in Fig.\ \ref{fig:closedAC}.
To obtain explicit results, we again consider a simplified model for the reflection and transmission from the magnetic insulator,
\begin{align}
  r_{\rm M}(\varepsilon) =&\, e^{2 i k_{\rm F} L_3} \theta(\Delta - |\varepsilon|), \nonumber \\
  r_{\rm M}'(\varepsilon) =&\, \theta(\Delta - |\varepsilon|), \nonumber \\
t_{\rm M}(\varepsilon) =&\, t_{\rm M}'(\varepsilon) = i e^{i k_{\rm F} L_3} \theta(|\varepsilon| - \Delta).
  \label{eq:simplemodel}
\end{align}
In comparison to the model, Eq.\ (\ref{eq:rMopen}), used in the previous section, we have made the further simplification that $\omega_{\rm M} L_3/v_{\rm F}$, $\omega_{\rm M} L/v_{\rm F} \ll 1$ already at the beginning of the calculation, which allows us to replace $k(\varepsilon)$ by $k_{\rm F}$ in the expression for $r_{\rm M}(\varepsilon)$ and in the phase factor $e^{i k(\varepsilon) L}$ for transmission through the interferometer arm. The phase shift associated with $r_{\rm M}'(\varepsilon)$ is absorbed into the phase factor $e^{i k L}$, i.e., the length $L$ of the upper arm is extended to include the distance between the (closed) right point contact and the magnet.
The correction to the DC current, Eq.\ (\ref{eq:IDCclosed}), vanishes in the simple model because the product $r_{\rm M}(\varepsilon) t_{\rm M}(\varepsilon) = r_{\rm M}(\varepsilon) t_{\rm M}'(\varepsilon) = 0$ at all energies $\varepsilon$. To keep the final expressions for the AC current components simple, we also set
the temperature to zero, and choose $V_1 = V$ and $V_2=V_7=V_8=0$. Solving Eq. (\ref{eq:omegaMclosed}) for the precession frequency $\omega_{\rm M}$ gives
\begin{align}
\hbar \omega_{\rm M} = - {\rm min}[eV, 2 \Delta] .
\end{align}
The AC components at $\omega=\omega_{\rm M}$ are then given by
\begin{widetext}
\begin{align}
  I_1(\omega_{\rm M}) =&\, \hat{m}_- \sqrt{T_1'} \frac{e}{h} e^{i \phi+i k_F
(L-L_3)} {\rm min}[ \Delta +3\hbar \omega_{\rm M}/2, 0] \nonumber \\ 
&\,- i \hat{m}_- \sqrt{T_1' R_1}\frac{e}{h} \left\{ e^{-2i k_F L_3} {\rm
min}[-\hbar \omega_{\rm M},\Delta+\hbar\omega_{\rm M}/2] + e^{2i k_F L}
\hbar\omega_{\rm M} \right\} ,\nonumber\\
I_2(\omega_{\rm M}) =&\, -i \hat{m}_- \sqrt{T_1' R_1}\frac{e}{h}
\left( e^{2 i k_F
L} - e^{-2 i k_F L_3} \right) {\rm min}[-\hbar \omega_{\rm M}, \Delta +
  \hbar\omega_{\rm M}/2] .
\end{align}
The AC components at $\omega = 2\omega_{\rm M}$ are
\begin{align}
I_1(2\omega_{\rm M}) = &\, -I_2(2\omega_{\rm M}) \nonumber\\
=&\, -i \hat{m}^2_- \frac{e}{h} T_1' e^{-2 i k_F(L-L_3) }
{\rm min}[-\hbar
\omega_{\rm M}, \Delta + \hbar\omega_{\rm M}/2]  .
\end{align}
The remarks from Sec.\ \ref{ssec:cl-vs-qu} apply when interpreting these AC current contributions in the case of a ``quantum'' magnet with sharply defined $M_z$.

\section{Full interferometer}\label{sec:full}

In this section we will now present results for the full interferometer (see Fig.\ \ref{fig:setup}), for which all reflection and transmission coefficients $R_j$, $T_j$, and $T_j'$, $j=1,2$, are nonzero. To keep the expressions concise, we give analytical results to lowest nontrivial order in $R_j$, $T_j'$, $j=1,2$, only. These are compared with a numerical evaluation of the full expression in Fig.\ \ref{fig:numcurrents}.

As before, the precession frequency of the magnet is obtained from the relation $\dot M_z = 0$. To first nonzero order on either side of the magnet we obtain
\begin{align}
  \label{eq:omegaMfullapprox}
  0 =&\,
  \int d\varepsilon \left\{ |r_{\rm M}(\varepsilon)|^2
  [f_2(\varepsilon_+) - f_7(\varepsilon_-)] +
  T_1' [f_1(\varepsilon_+) - f_2(\varepsilon_+)]
  + R_1 [f_8(\varepsilon_+) - f_2(\varepsilon_+)]
  \right. \nonumber \\ &\, \vphantom{\int} \left. \ \ \ \ \mbox{}
  + T_2' [f_7(\varepsilon_-) - f_8(\varepsilon_-)]
  + R_2 [f_7(\varepsilon_-) - f_1(\varepsilon_-)] \right\},
\end{align}
where we also kept subleading terms because the leading contribution may vanish if $V_2 = V_7$. The leading contribution to the DC currents in the four leads is then
\begin{align}
\label{eq:DCfullapprox}
  I_1(0)=&\, \frac{e}{h} \int d\varepsilon \left\{ f_1(\varepsilon) - f_8(\varepsilon) \right\}
    \nonumber \\ &\, \mbox{}
       + \frac{2 e}{h}\, \mbox{Re}\, \int d\varepsilon \left\{
       \sqrt{R_1 R_2} t_{\rm M}(\varepsilon_-) e^{-i(\phi - k L)}
         [f_2(\varepsilon)-f_8(\varepsilon)]
       \mbox{}
       + \sqrt{T_1' T_2'} t_{\rm M}'(\varepsilon_+) e^{i(\phi - k L)}
         [f_8(\varepsilon)-f_7(\varepsilon)]
       \right\} , \nonumber \\
  I_2(0)=&\, \frac{e}{h} \int d\varepsilon \left\{ f_2(\varepsilon) - f_7(\varepsilon) + |r_{\rm M}(\varepsilon_-)|^2
       [ f_7(\varepsilon-\hbar \omega_{\rm M}) - f_2(\varepsilon) ] \right\}
    \nonumber \\ &\, \mbox{}
       + \frac{2 e}{h}\, \mbox{Re}\, \int d\varepsilon \left\{
       \sqrt{R_1 R_2} t_{\rm M}'(\varepsilon_+) e^{i(\phi + k L)} [f_1(\varepsilon)-f_7(\varepsilon)]
       + \sqrt{T_1' T_2'} t_{\rm M}'(\varepsilon_+) e^{i(\phi - k L)} [f_7(\varepsilon)-f_8(\varepsilon)]
       \right. \nonumber \\ &\, \ \ \ \ \left. \mbox{}
       + \sqrt{R_1 R_2} |r_{\rm M}(\varepsilon_-)|^2
       [ t_{\rm M}(\varepsilon_-) e^{-i(\phi - k L)} + t_{\rm M}'(\varepsilon_-) e^{i(\phi + k_- L)}]
       [f_7(\varepsilon-\hbar \omega_{\rm M}) - f_2(\varepsilon)] \right\}, \nonumber \\
  I_7(0)=&\, \frac{e}{h} \int d\varepsilon \left\{ f_7(\varepsilon) - f_2(\varepsilon) - |r_{\rm M}(\varepsilon_-)|^2
  [ f_7(\varepsilon-\hbar \omega_{\rm M}) - f_2(\varepsilon) ] \right\}
       \nonumber \\ &\, \mbox{}
       + \frac{2e}{h}\, \mbox{Re}\, \int d\varepsilon
       \left\{ \sqrt{R_1 R_2} t_{\rm M}(\varepsilon_-) e^{-i(\phi - k L)}
       [f_8(\varepsilon)-f_2(\varepsilon)]
       + \sqrt{T_1' T_2'} t_{\rm M}(\varepsilon_-) e^{-i(\phi + k L)}
       [f_2(\varepsilon)-f_1(\varepsilon)]
       \right. \nonumber \\ &\, \ \ \ \ \left. \mbox{}
       + \sqrt{R_1 R_2} |r_{\rm M}(\varepsilon_-)|^2
       [t_{\rm M}(\varepsilon_-) e^{-i(\phi - k L)} +
         t_{\rm M}'(\varepsilon_-) e^{i(\phi + k_- L)}]
       [f_2(\varepsilon)-f_7(\varepsilon-\hbar \omega_{\rm M})] \right\}, \nonumber \\
  I_8(0)=&\, \frac{e}{h} \int d\varepsilon \left\{ f_8(\varepsilon) - f_1(\varepsilon) \right\}
       \nonumber \\ &\, \mbox{}
       + \frac{2e}{h}\, \mbox{Re}\, \int d\varepsilon \left\{
       \sqrt{R_1 R_2} t_{\rm M}'(\varepsilon_+) e^{i(\phi + k L)}
       [f_7(\varepsilon)-f_1(\varepsilon)]
       + \sqrt{T_1' T_2'} t_{\rm M}(\varepsilon_-) e^{-i(\phi + k L)}
       [f_1(\varepsilon)-f_2(\varepsilon)] \right\},
\end{align}
where the subleading terms contain the $\phi$-dependent contribution to the DC current only. Similarly, for the AC currents we find
\begin{align}
\label{eq:ACfullapprox}
I_1(\omega_{\rm M})=&\, i \hat{m}_- \frac{e}{h} \int d\varepsilon \left\{ \sqrt{R_2 T_2'} r_{\rm M}'(\varepsilon_+) e^{-i L (k-k_+)} \left[ f_7(\varepsilon)-f_8(\varepsilon) \right] + \sqrt{R_1 T_1'} r_{\rm M}^*(\varepsilon_-) \left[ f_8(\varepsilon) - f_2(\varepsilon) \right] \right\} , \nonumber \\
I_2(\omega_{\rm M})=&\, i \hat{m}_- \sqrt{R_1 T_1'} \frac{e}{h} \int d\varepsilon r_{\rm M}^*(\varepsilon_-) \left[ f_1(\varepsilon)-f_8(\varepsilon) \right] , \nonumber \\
I_7(\omega_{\rm M})=&\, -i \hat{m}_- \sqrt{R_2 T_2'} \frac{e}{h} \int d\varepsilon r_{\rm M}'(\varepsilon_+) \left[ f_1(\varepsilon)-f_8(\varepsilon) \right] , \nonumber \\
I_8(\omega_{\rm M})=&\, i \hat{m}_- \frac{e}{h} \int d\varepsilon \left\{ \sqrt{R_2 T_2'} r_{\rm M}'(\varepsilon_+) \left[ f_1(\varepsilon)-f_7(\varepsilon) \right] + \sqrt{R_1 T_1'} r_{\rm M}^*(\varepsilon_-) e^{-i L (k_- - k)} \left[ f_2(\varepsilon) - f_1(\varepsilon) \right] \right\} ,\\
I_1(2\omega_{\rm M})=&\, - I_2(2 \omega_{\rm M}) \nonumber \\ =&\,
- \hat{m}_-^2 \frac{e}{h} \int d\varepsilon e^{-iL(k - k_+)} r_{\rm M}'(\varepsilon_+) r_{\rm M}^*(\varepsilon_-) 
\left\{ R_2 T_1' e^{2iLk} [f_1(\varepsilon) - f_2(\varepsilon) ] + \sqrt{R_1 R_2 T_1' T_2'} [f_8(\varepsilon) - f_7(\varepsilon)] \right\} , \nonumber \\
I_7(2\omega_{\rm M})=&\, -I_8(2 \omega_{\rm M}) \nonumber \\ =&\,
\hat{m}_-^2 \frac{e}{h} \int d\varepsilon e^{-iL(k_- + k)} r_{\rm M}'(\varepsilon_+) r_{\rm M}^*(\varepsilon_-) 
  \left\{ \sqrt{R_1 R_2 T_1' T_2'} e^{2iLk} [f_1(\varepsilon) - f_2(\varepsilon)] + R_1 T_2' [f_8(\varepsilon) - f_7(\varepsilon)] \right\}.
  \label{eq:AC2fullapprox}
\end{align}
\end{widetext}

For the simplified model defined by Eq.\ (\ref{eq:simplemodel}) and the subsequent discussion we compare the expressions (\ref{eq:omegaMfullapprox})--(\ref{eq:AC2fullapprox}) for small $T_j'$, $R_j$, $j=1,2$, with a numerical evaluation of the full solution of the scattering problem. Figure \ref{fig:numcurrents} shows the result of this comparison for the bias voltages $V_1=V>0$, $V_2=V_7=V_8=0$. The $\phi$-dependent contribution to the DC current in lead ``2'' shown in Fig. \ref{fig:numcurrents} is calculated as
\begin{align}
	\label{eq:ABstandarddev}
	I_{2,AB}(0)\equiv \frac{1}{2 \pi}\left\{ \int_0^{2\pi} d\phi \left[ I_2(0)^2 - \langle I_2(0) \rangle^2 \right] \right\}^{1/2},
\end{align}
where
\begin{align}
	\langle I_2(0) \rangle = \frac{1}{2\pi}\int_0^{2\pi}d\phi I_2(0).
\end{align}
\begin{figure}
	\centering
	\includegraphics[width=.45\textwidth]{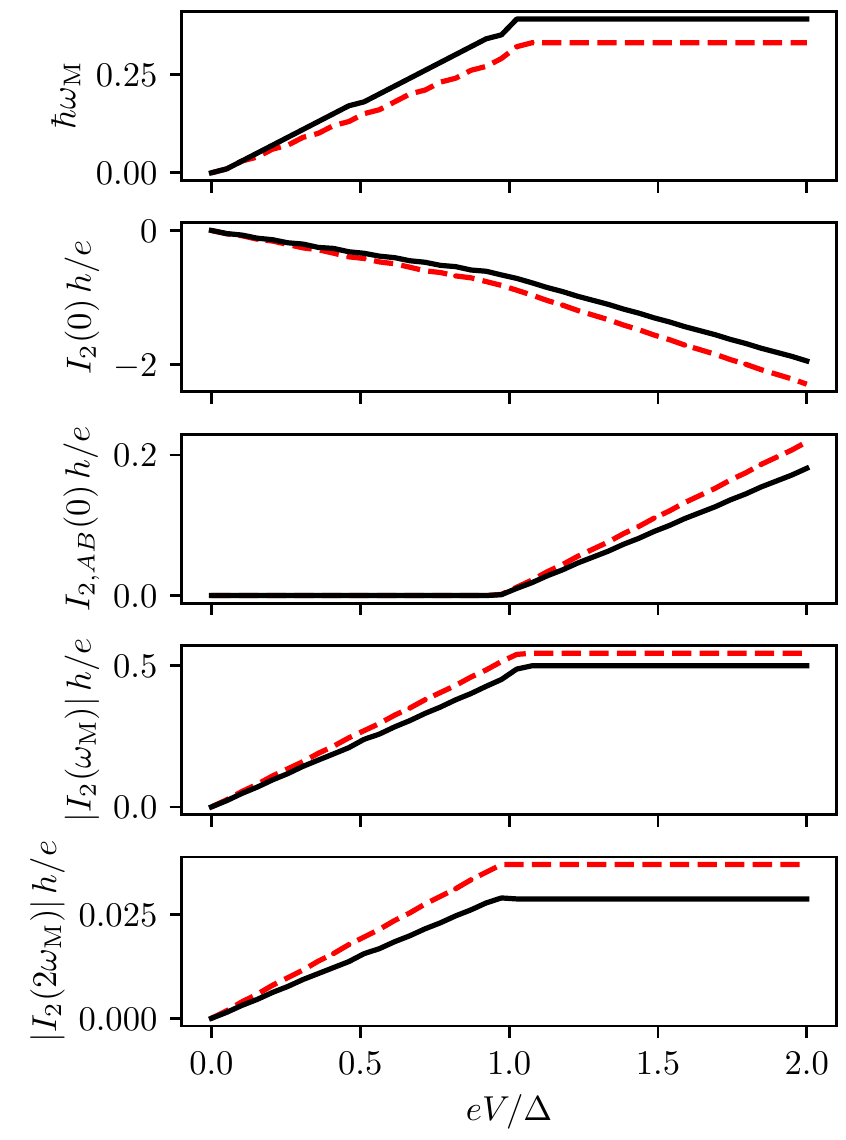}
	\caption[]{From top to bottom: Precession frequency $\omega_{\rm M}$, DC current $I_2(0)$ at zero Aharonov-Bohm phase $\phi$, $\phi$-dependent contribution to DC current defined in Eq. (\ref{eq:ABstandarddev}),
and the magnitude $|I_2(\omega_{\rm M})|$ and $|I_2(2\omega_{\rm M})|$ of the AC current components at frequency $\omega_{\rm M}$ and $2\omega_{\rm M}$, as a function of the bias voltage $V$ at lead ``1.'' All currents are evaluated in lead ``2.'' The solid curves show the numerical evaluation of the full solution without the approximation of small $R_j$, $T_j'$, $j=1,2$. The dashed curves are obtained using the analytical expressions (\ref{eq:omegaMfullapprox})--(\ref{eq:AC2fullapprox}), with the exception of the second panel, where we have included all $\phi$-independent terms up to second order in  $(R_j)^{1/2}$, $(T_j')^{1/2}$, not only those of lowest nontrivial order given in $I_2(0)$ in Eqs.\ (\ref{eq:DCfullapprox}). The parameters are chosen as $T_1=0.79$, $R_1=0.09$, $T_1'=0.12$, $T_2=0.78$, $R_2=0.06$, $T_1'=0.16$, $k_F L =1$, and $k_F L_3=0.1$.}
	\label{fig:numcurrents}
\end{figure}

\section{Finite-$T$ enhancement of DC Aharonov-Bohm currents}\label{sec:finiteTAB}

All previous explicit results for the simple model, Eq.\ (\ref{eq:simplemodel}), were calculated at zero temperature ($T=0$). In this section we consider the effect of the Aharonov-Bohm phase $\phi$ on the DC current at finite temperature. We find that, if the Fermi energy is in the magnet-induced gap, increasing $T$ leads to an increase of the $\phi$-dependent Aharonov-Bohm contributions to the DC current. The reason is that coherent transmission through the magnet, which is required for a dependence on the Aharanov-Bohm phase $\phi$, exists for above-gap energies only. The population of the above-gap states increases with temperature, which causes the $\phi$-dependent current contribution to increase. The visibility of the Aharonov-Bohm oscillations decreases again at higher temperatures, when thermal smearing effects start to dominate.
\begin{figure}[h!]
	\centering
	\includegraphics[width=.45\textwidth]{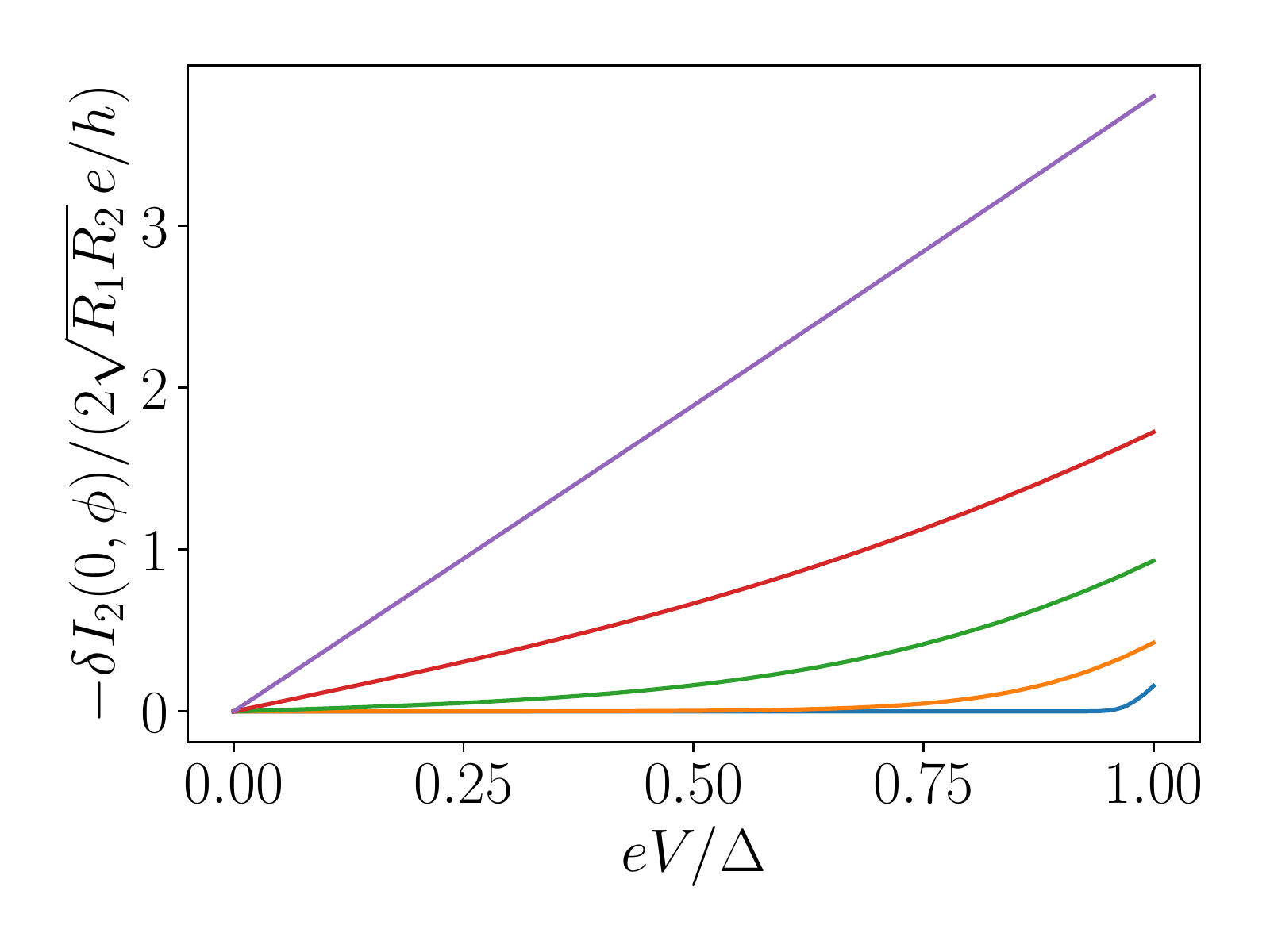}
	\caption{Aharonov-Bohm current $\delta I_2(0,\phi)$ from Eq. (\ref{eq:TdepABfullapprox2}) as a function of applied bias $V$ in lead ``1'' for several values of $k_{\rm B}T/\Delta$; from bottom to top they are $0.01,\,0.1,\,0.25,\,0.5$ and $2$. The precession frequency $\omega_{\rm M}$ is determined from the solution of Eq. (\ref{eq:omegaMfullapprox}) at finite $T$. The phases are chosen such that $\sin[k_F(L_3+L)+\phi]=1$.}
	\label{fig:TdepABfull}
\end{figure}
As before, we choose the biases $V_1=V>0$ and $V_2=V_7=V_8=0$ and use the simple model given in Eq.\ (\ref{eq:simplemodel}) to calculate the $\phi$-dependent contributions to the DC current. At finite temperature, the distribution functions are $f_j(\varepsilon) = [1 + e^{(\varepsilon - e V_j)/k_{\rm B} T}]^{-1}$ with equal temperature $T$ applied to all four leads $j=1,2,7,8$.

Flux-dependent contributions to the DC current do not appear in the ``open'' geometry of Sec.\ \ref{sec:open}. They do appear in principle for the ``closed'' geometry of Sec.\ \ref{sec:closed}, but not for the simple model given in Eq.\ (\ref{eq:simplemodel}), since all the $\phi$-dependent terms of Eq. (\ref{eq:IDCclosed}) are proportional to $r_{\rm M}(\varepsilon_{-}) t_{\rm M}(\varepsilon_{-})$, which is zero for that model. Hence, the discussion here focuses on the full interferometer geometry of Sec.\ \ref{sec:full}. The starting point of our analysis is Eq.\ (\ref{eq:DCfullapprox}), which gives the DC currents $I_j(0)$ for general distribution functions $f_j(\varepsilon)$. Specializing to the simple model, Eq.\ (\ref{eq:simplemodel}), and restricting to the flux-dependent part $\delta I_j(0,\phi)$ of the DC current, we find
\begin{widetext}
\begin{align}
\label{eq:TdepABfullapprox}
\delta I_1(0,\phi)=&\, 0 , \nonumber \\
\delta I_2(0,\phi)=&\, - \frac{2e}{h} \sqrt{R_1 R_2} \sin{[k_F (L_3 + L) + \phi]}
\left\{ \int_{-\infty}^{-\Delta-\hbar \omega_{\rm M}/2} d\varepsilon [f_1(\varepsilon)-f_7(\varepsilon)]
+ \int_{\Delta-\hbar \omega_{\rm M}/2}^{\infty} d\varepsilon [f_1(\varepsilon)-f_7(\varepsilon)] \right\} , \nonumber \\
\delta I_7(0,\phi)=&\, - \frac{2e}{h} \sqrt{T_1' T_2'} \sin{[k_F (L_3 - L) - \phi]}
\left\{ \int_{-\infty}^{-\Delta+\hbar \omega_{\rm M}/2} d\varepsilon [f_2(\varepsilon)-f_1(\varepsilon)]
+ \int_{\Delta+\hbar \omega_{\rm M}/2}^{\infty} d\varepsilon [f_2(\varepsilon)-f_1(\varepsilon)] \right\} , \nonumber \\
\delta I_8(0,\phi)=&\, -I_2(0,\phi)-I_7(0,\phi).
\end{align}
The integrals can be performed analytically and for our choice of biases they are
\begin{align}
\label{eq:TdepABfullapprox2}
  \delta I_2(0,\phi)=&\, - \frac{2e k_{\rm B}T}{h} \sqrt{R_1 R_2} \sin{[k_F (L_3 + L) + \phi]}
\left[ \frac{eV}{k_{\rm B}T} + {\rm ln}\, {\rm cosh} \frac{\Delta+\hbar \omega_{\rm M}/2}{2 k_{\rm B}T} - {\rm ln}\, {\rm cosh} \frac{e V+\Delta+\hbar \omega_{\rm M}/2}{2 k_{\rm B}T} \right. \nonumber \\
&\left. \ \ \mbox{} - {\rm ln}\, {\rm cosh} \frac{\Delta-\hbar \omega_{\rm M}/2}{2 k_{\rm B}T} + {\rm ln}\, {\rm cosh} \frac{e V - \Delta+\hbar \omega_{\rm M}/2}{2 k_{\rm B}T}  \right] , \nonumber \\
 \delta I_7(0,\phi)=&\, \frac{2e k_{\rm B}T}{h} \sqrt{T_1' T_2'} \sin{[k_F (L_3 - L) - \phi]}
\left[ \frac{eV}{k_{\rm B}T} + {\rm ln}\, {\rm cosh} \frac{\Delta-\hbar \omega_{\rm M}/2}{2 k_{\rm B}T} - {\rm ln}\, {\rm cosh} \frac{e V +\Delta-\hbar \omega_{\rm M}/2}{2 k_{\rm B}T} \right. \nonumber \\
&\left. \ \ \mbox{} - {\rm ln}\, {\rm cosh} \frac{\Delta+\hbar \omega_{\rm M}/2}{2 k_{\rm B}T} + {\rm ln}\, {\rm cosh} \frac{e V - \Delta-\hbar \omega_{\rm M}/2}{2 k_{\rm B}T}  \right] , \nonumber \\ .
\end{align}
\end{widetext}
In the limit of $k_B T \gg \Delta,\,eV,\,\hbar \omega_{\rm M}$, these expressions reduce to
\begin{align}
  \delta I_2(0,\phi) =&\, - V \frac{2e^2}{h} \sqrt{R_1 R_2} \sin{[k_F (L_3 + L) + \phi]} \nonumber ,\\
  \delta I_7(0,\phi) =&\,  V \frac{2e^2}{h} \sqrt{T_1' T_2'} \sin{[k_F (L_3 - L) - \phi]} ,
\end{align}
whereas in the opposite limit $k_BT\rightarrow 0$, we find that the currents vanish, consistent with our previous results. The existence of a temperature-independent visibility of the Aharonov-Bohm oscillations of the DC current at high temperatures continues up to temperatures $k_{\rm B} T \sim \min(L,L_3)/\hbar v_{\rm F}$ where $L$ and $L_3$ are the lengths of the segments of the Aharonov-Bohm ring. The visibility of the Aharonov-Bohm effect is suppressed at higher temperatures because of thermal smearing effects [not present in the simple model (\ref{eq:simplemodel})].
In Fig. \ref{fig:TdepABfull}, we show $\delta I_2(0,\phi)$ from Eq. (\ref{eq:TdepABfullapprox2}), confirming the enhancement of the Aharonov-Bohm oscillations compared to the zero-temperature result of panel 3 of Fig.\ \ref{fig:numcurrents}.

\section{Conclusion} \label{sec:conclusion}

In this work, we analyzed electrical transport through a helical edge of a two-dimensional topological insulator exchange-coupled to a magnetic insulator. Despite the presence of an excitation gap, the magnet has no effect on the current if it has an easy-plane anisotropy with the easy plane perpendicular to the spin quantization axis of the helical edge.\cite{meng2014,silvestrov2016} Here, we show that the exchange coupling to the magnet does affect electrical transport in an interferometer geometry: (1) In a four-terminal geometry, the application of a DC voltage leads to AC currents at frequencies $\omega_{\rm M}$ and  $2\omega_{\rm M}$ with $\omega_{\rm M}$ being the precession frequency of the magnet's magnetization and (2) if the Fermi energy is in the magnet-induced gap, the usual Aharonov-Bohm-flux dependent oscillations of the DC current are strongly suppressed at zero temperature or bias voltage and show a maximum at temperature or voltage comparable to the magnet-induced gap.

Time-reversal symmetry prohibits backscattering at the two point contacts that define the interferometer. In the limit of helical edges with a well-defined spin polarization, forward scattering between different edges is also suppressed (although it will still be finite), because it requires a spin flip. The alternating currents exist only if there is such spin-flip scattering at the point contacts. They can be traced back to an interference contribution between transmission paths through the interferometer that involve a (spin-flipping) reflection from the precessing magnet and a spin-flip scattering at the contacts. The precessing magnetization inserts a time-dependent phase factor in this interference contribution, thus causing an alternating current contribution for a time-independent applied bias. This origin of the alternating current contribution as a periodic modulation of the interference correction to the (steady-state) conductance must be contrasted with the origin of the direct current through the magnet, which can be seen as a current ``pumped'' by the precession magnetization.\cite{meng2014} Consequentially, the magnitude of the direct current through the magnet is closely related to the precession frequency---it is one electron per period of the precessing magnetization---whereas, depending on the properties of the point
contacts and the specific way of biasing the interferometer, the magnitude of the alternating current components can be smaller or larger than that.

An easy-plane ferromagnet exhibits a ``spin superfluid'' state,\cite{koenig2001,sonin2010} which has its origin in the formal analogy between the $U(1)$ low-energy degree of freedom of a magnetic moment with easy-plane anisotropy and the $U(1)$ freedom of the superfluid phase.\cite{nogueira2004} Under suitable external driving, such a spin superfluid can enter into a spiral state in which it carries a dissipationless spin current.\cite{koenig2001} In nontopological systems, the bottleneck for observing this dissipationless current is its conversion to measurable spin or change currents outside the magnet.\cite{takei2014,takei2015} The system that we consider here (and that was previously considered in Refs.\ \onlinecite{meng2014,silvestrov2016}) offers a scenario for a perfect conversion between the charge current in the helical edge of a two-dimensional topological insulator and the ``spin superfluid'' of the easy-plane ferromagnet. In addition to the absence of shot noise predicted in Ref.\ \onlinecite{silvestrov2016}, the transport properties we identify in this paper are unique signatures of the anomalous electric transport in this system.

\begin{acknowledgments}
P.R. thanks Adrian Schneider for insightful discussions in an early stage of the project.
We acknowledge financial support by the German Science Foundation (DFG) through Grants No. RE 2978/8-1 (P.S., P.R.) and No. BR 7074/4-1 (P.W.B.) and by project A02 of the CRC-TR 183 ``entangled states of matter'' (P.W.B., K.M.). P.R. and P.S. further acknowledge support from the DFG under Germany's Excellence Strategy, EXC-2123 QuantumFrontiers, 390837967.
\end{acknowledgments}

\bibliography{bibliography}

\end{document}